\documentclass[aps,pre,showpacs,showkeys,amsmath,amsfonts,amssymb]{revtex4} 
\usepackage[dvips]{graphicx}
\usepackage[latin1]{inputenc}
\usepackage{wrapfig}
\usepackage{psfrag}

\newcommand{\beq}{\begin{equation}} 
\newcommand{\eeq}{\end{equation}}
\newcommand{\bea}{\begin{eqnarray}} 
\newcommand{\eea}{\end{eqnarray}} 
 
\input epsf 
 
\begin{document} 
 
\title{Geometry of the energy landscape and folding transition in a simple model
of a protein} 
 
\author{Lorenzo N.\ Mazzoni} 
\email{mazzoni@fi.infn.it} 
\affiliation{Dipartimento di Fisica, Universit\`a di 
Firenze, via G.~Sansone 1, I-50019 Sesto Fiorentino (FI), Italy}  
\author{Lapo Casetti} 
\email{lapo.casetti@unifi.it} 
\affiliation{Dipartimento di Fisica and Centro per lo Studio
delle Dinamiche Complesse (CSDC), Universit\`a di 
Firenze, and Istituto Nazionale di Fisica Nucleare (INFN), 
Sezione di Firenze, via G.~Sansone 1, I-50019 Sesto Fiorentino (FI), Italy}  

\date{\today} 
 
\begin{abstract} 
A geometric analysis of the global properties of the energy landscape of a minimalistic
model of a polypeptide is presented, which is based on the relation between dynamical
trajectories and geodesics of a suitable manifold, whose metric is completely determined
by the potential energy. We consider different sequences, some with a definite
protein-like behavior, a unique native state and a folding transition, and the others
undergoing a hydrophobic collapse with no tendency to a unique native state. The global
geometry of the energy landscape appears to contain relevant information on the behavior
of the various sequences: in particular, the fluctuations of the curvature of the energy
landscape, measured by means of numerical simulations, clearly mark the folding
transition and allow to distinguish the protein-like sequences from the others. 
\end{abstract} 
 
\pacs{87.15.-v; 02.40.-k} 
 
\keywords{Polymers; protein folding; geometry; energy landscape} 
 
\maketitle 

\section{Introduction}

Protein folding is one of the most fundamental and challenging open questions
in molecular biology. Proteins are polymers made of amino acids and since the
pioneering experiments by Anfinsen and coworkers \cite{Anfinsen} it has been
known that the sequence of amino acids---also called the primary structure of
the protein---uniquely determines its native state, or tertiary structure,
i.e., the compact configuration the protein assumes in physiological conditions
and which makes it able to perform its biological tasks \cite{proteinbook}. To
understand how the information contained in the sequence is translated into the
three-dimensional native structure is the core of the protein folding problem,
and its solution would allow one to predict a protein's structure from the sole
knowledge of the amino acid sequence: being the sequencing of a protein
much easier than experimental determination of its structure, it is easy to understand
the impact such a solution would have on biochemistry and molecular biology. 
Moreover, solving the protein folding
problem would make it possible to engineer proteins which fold to any given
structure (what is commonly referred to as the inverse folding problem), which
in turn would mean a giant leap in drug design. Despite many remarkable
advances in the last decades, the protein folding problem is
still far from a solution \cite{proteinbook}.

A polymer made of amino acids is referred to as a polypeptide. However,
not all polypeptides are proteins: only a very small subset of all the possible
sequences of the twenty naturally occurring amino acids have been selected by
evolution. According to our present knowledge, all the naturally selected
proteins fold to a uniquely determined native state, but a generic polypeptide
does not (it rather has a glassy-like behavior). Then the following question
naturally arises: what makes a protein different from a generic polypeptide?
or, more precisely, which are the properties a polypeptide must have to behave
like a protein, i.e., to fold into a unique native state regardless of the
initial conditions, when the environment is the correct one?
This question is, obviously, much less general than the whole folding problem,
nonetheless if one could give it an answer it would surely help in the quest of
a solution to the folding problem. 

This question has aroused a lot of interest in the recent years:
the {\em energy landscape} picture has emerged as crucial in this respect.
Energy landscape, or more precisely potential energy landscape, is the name
commonly given to the graph of the potential energy of interaction between the
microscopic degrees of freedom of the system \cite{Wales}; the
latter is a high-dimensional surface, but one can also speak of a free energy
landscape when only its projection on a small set of collective variables (with
a suitable average over all the other degrees of freedom) is considered
\cite{Wales}. Before having been applied to biomolecules, this concept has
proven useful in the study of other complex systems, especially of supercooled
liquids and of the glass transition \cite{naturecomplex}. The basic idea is
very simple, yet powerful: if a system has a rugged, complex energy landscape,
with many minima and valleys separated by barriers of different height, its
dynamics will experience a variety of time scales, with oscillations in the
valleys and jumps from one valley to another\footnote{For a
classical dynamics at a finite temperature, with a deterministic force given by 
the gradient of the potential energy and a stochastic
force proportional to the temperature, responsible for the thermally activated
jumps between valleys. However, for a sufficiently large system also a
completely deterministic dynamics over the same landscape would give a similar
overall behavior}. Then one can try to link special features of the behavior of
the system (i.e., the presence of a glass transition, the separation of time
scales, and so on) to special properties of the landscape, like the topography
of the basins around minima, the energy distribution of minima and saddles
connecting them and so on. Anyway, a complex landscape yields a complex
dynamics, where the system is very likely to remain trapped in different
valleys when the temperature is not so high. This is consistent with a glassy
behavior, but a protein does not show a glassy behavior, it rather has
relatively low frustration. This means that there must be some property of the
landscape such to avoid too much frustration. This property is commonly
referred to as the {\em folding funnel} \cite{funnel}: though locally rugged,
the low-energy part of the energy landscape is supposed to have an overall
funnel shape so that most initial conditions are driven towards the correct
native state. The dynamics must then be such as to make this happen in a reasonably fast
and reliable way, i.e., non-native minima must be efficiently connected to the native
state si that trapping in the wrong onfiguration si unlikely.

However, a direct visualization of the energy landscape is
impossible due to its high dimensionality, and its detailed properties must be
inferred indirectly. A possible strategy is a local one: one searches for the
minima of the landscape and then for the saddles connecting different minima.
This is practically unfeasible for accurate all-atom potential energies, but
may become accessible for minimalistic potentials\footnote{Searching for {\em
all} the minima and {\em all} the saddles is impossibile even for very simple
potentials, unless it can be done analytically, because no algorithm is
available which is able to find {\em all} the solutions of $N$ coupled
nonlinear equations \protect\cite{numrec}; nonetheless, one expects that with a
considerable numerical effort a reasonable sampling of these points can be
achieved for not-too-detailed potentials, as it happens for binary
Lennard-Jones fluids \protect\cite{Grigera}.}. Minimalistic models are those
where the polymer is described at a coarse-grained level, as a chain of $N$
beads where $N$ is the number of amino acids; no explicit water molecules are
considered and the solvent is taken into account only by means of effective
interactions among the monomers. Minimalistic models can be relatively simple,
yet in some cases yield very accurate results which compare well with
experiments \cite{Clementi_review,Clementi}. The local properties of the energy landscape of
minimalistic models have been recently studied 
(see e.g.\ Refs.\ \cite{Miller_Wales,BonginiRampioni,Bongini1,keyes,peyrard,Luccioli}) 
and very interesting clues about
the structure of the folding funnel and the differences between protein-like
heteropolymers and other polymers have been found: in particular, it has 
been shown that a funnel-like structure is present also in homopolymers, 
but what makes a big
difference is that in protein-like systems jumps between minima corresponding
to distant configurations are much more favoured dynamically \cite{Bongini2}.

The above mentioned local strategy to analyze energy landscapes requires
however a huge computational effort if one wants to obtain a good sampling. So
the following question arises: is there some {\em global} property of
the energy landscape which can be easily computed numerically as an average
along dynamical trajectories and which is able to identify polymers having a
protein-like behavior? We shall show in the following that
such a quantity indeed exists, at least for the minimalistic model we
considered, and that it is of a geometric nature. In particular, 
the fluctuations of a suitably defined curvature of the energy landscape
clearly mark the folding transition while do not show any remarkable feature
when the polymer undergoes a hydrophobic collapse without a preferred native
state. This is at variance with thermodynamic global observables, like the
specific heat, which show a very similar behavior in the case of a folding
transition and of a simple hydrophobic collapse. 

The paper is organized as follows: we first describe the geometric approach to
energy landscapes, then we discuss the model studied and our results.
A final section is devoted to some comments. 
A short, preliminary account of a part of the results presented here has already been given in
\cite{mazzoni_casetti}.

\section{Geometry of the energy landscape}

The intuitive reason why geometric information on the landscape, and especially
curvature, could be relevant to the problem of folding is that the dynamics on
a landscape would be heavily affected by the local curvature: minima of the
energy landscape are associated to positive curvatures and stable dynamics,
while saddles involve negative curvatures, at least along some direction, thus
implying some instability. One can reasonably expect that the arrangement and
detailed properties of minima and saddles might reflect in some global feature
of the distribution of curvatures of the landscape, when averaged along a
typical trajectory.

The  definition of the
curvature of a manifold $M$ depends on the choice of a metric $g$
\cite{Nakahara,DoCarmo}: once the couple $(M,g)$ is given, a covariant
derivative and a curvature tensor $R(e_i,e_j)$ can be defined; the latter
measures the noncommutativity of the covariant derivatives in the coordinate
directions $e_i$ and $e_j$. A scalar measure of the curvature at any given
point $P \in M$  is the  the sectional curvature 
\beq
K(e_i,e_j) = \langle R(e_i,e_j) e_j,e_i\rangle~,
\eeq
where $\langle \cdot,\cdot \rangle$ stands for the
scalar product. At any point of an $N$-dimensional manifold there are $N(N-1)$
sectional curvatures, whose knowledge determines the full curvature tensor at
that point. One can however define some  simpler
curvatures (paying the price of losing some information): 
the Ricci curvature $K_R(e_i)$ is the sum of the $K$'s over the $N-1$
directions orthogonal to $e_i$, 
\beq
K_R(e_i) = \sum_{j = 1}^N
K(e_i,e_j)~, \label{riccicurv}
\eeq
and summing also on the $N$ directions $e_i$ one gets the scalar
curvature 
\beq
{\cal R} = \sum_{i = 1}^N K_R(e_i) =
\sum_{i,j = 1}^N K(e_i,e_j)~;
\eeq  
then, $\frac{K_R}{N-1}$ and $\frac{{\cal R}}{N(N-1)}$ can be considered as
average curvatures at a given point.

Although one expects the association between minima and positive curvatures on
the one side and negative curvatures along some directions and saddles on the
other side to be essentially true for most choices of the metric, a particular
choice of $g$ among the many possible ones must be made in order to perform
explicit calculations. The most immediate choice would probably be that of
considering as our manifold $M$ the $N$-dimensional surface $z =
V(q_1,\ldots,q_N)$ itself, i.e., the graph of the potential energy $V$ as a
function of the $N$ coordinates $q_1,\ldots,q_N$ of the configuration space,
and to define $g$ as the metric induced on that surface by its immersion in
$\mathbb{R}^{N+1}$. Although perfectly reasonable, this choice has two
drawbacks: $(i)$ the explicit expressions for the curvatures in terms of
derivatives of $V$ are rather complicated and $(ii)$ the link between the
properties of the dynamics and the geometry is not very precise, i.e., one cannot
prove that the geometry completely determines the dynamics and its stability. 
For these reasons
we left the investigation of this particular geometry to future work and we
considered a choice of $(M,g)$ such that the link between geometry and dynamics
is more clear.

Let us now describe this metric and its relation to dynamics. We shall mention
only the most important results, referring the reader to the review paper
\cite{physrep} for the details. 

\subsection{Geometry and dynamics}

Let us consider a standard Hamiltonian system, with Hamiltonian function 
of the form
\begin{equation}
H=\sum_{i=1}^{N} \frac{p_{i}^{2}}{2m_{i}}+V(q_{1},...,q_{N}), \label{1_mio}
\end{equation} 
where $q_{i}$ and $p_{i}$ are the canonically conjugated coordinates and
momenta of the system, $N$ is the number of degrees of freedom and $m_{i}$ are the masses; 
in the following we shall consider $m_{i}=1$ $\forall i$. 

The trajectories of the system (\ref{1_mio}) in configuration space can be seen as
geodesics of a Riemannian manifold: this classic result is based on the variational
formulation of dynamics \cite{Arnold}. Hamilton's principle states that the natural motions of a
system are the extrema of the action functional
\begin{equation}
S =\int L \ dt \ ,  \label{2_mio}
\end{equation} 
where $L$ is the Lagrangian function of the system,
\begin{equation}
L=\frac{1}{2}\delta_{ij}\dot{q}^{i}\dot{q}^{j}-V(q_{1},...,q_{N})  \label{5_mio}
\end{equation} 
(summation on repeated indices is understood from now on); 
on the other hand, the geodesics of a Riemannian manifold are extrema of the length 
functional
\begin{equation}
\ell=\int ds \ , \label{4_mio}
\end{equation} 
where $s$ is the arc-length defined by
\beq
ds^2 = g_{ij}\, dq^i\, dq^j~,
\eeq
so that we can identify the geodesics with the physical trajectories of the system
by choosing a suitable metric $g$ linking action and length.

The typical example of such a metric 
is the Jacobi metric on the $N$-dimensional configuration space $M$, 
\begin{equation}
g_{ij}=2\left[E-V(q_{1},...,q_{N})\right]\, \delta_{ij} ~, \label{8_mio} 
\end{equation}
where $E$ is the total energy of the system. Starting from Eqs.\ (\ref{8_mio}) one can
easily show that the geodesic equations, i.e., the Euler-Lagrange equations for the
length functional (\ref{4_mio}), whose expression in local coordinates is  
\begin{equation}
\frac{d^{2}q^{i}}{ds^{2}}+\Gamma_{jk}^{i}\frac{dq^{j}}{ds}\frac{dq^{k}}{ds}=0 \ , \label{9_mio}
\end{equation} 
where the $\Gamma$'s are the Christoffel symbols\footnote{The expression of the
Christoffel symbols in local coordinates is $\Gamma^i_{jk} = \frac{1}{2}{g}^{im}  
\left( \partial_j {g}_{km} + \partial_k {g}_{mj} - \partial_m 
{g}_{jk} \right)$, where $\partial_i \equiv \partial/\partial q_i$ \protect\cite{Nakahara,DoCarmo}.}, 
become Newton's equations
\begin{equation}
\frac{d^{2}q^{i}}{dt^{2}}=-\frac{\partial V}{\partial q_{i}} \ . \label{10_mio}
\end{equation} 

However, the choice of such a metric is not unique. A particularly useful metric was
introduced by Eisenhart in 1929 \cite{Eisenhart}; it is the one we used and we are going
to describe in the following.

\subsubsection{Eisenhart metric and landscape curvature}

One could be tempted to identify actions with lengths by considering the configuration {\em
spacetime}, i.e., $M\times \mathbb{R}$, with local coordinates $(q_{0},q_{1},...,q_{N})$
where $q_0$ is the time coordinate, however it can't be easily done. Eisenhart's idea was
to further enlarge the configuration spacetime with another dimension, i.e., to consider
$M\times \mathbb{R}^2$ with local coordinates $(q_{0},q_{1},...,q_{N},q_{N+1})$, and to
endow this manifold with a pseudo-Riemannian metric whose arc-length is
\begin{equation}
ds^{2}=\delta_{i,j}dq^{i}dq^{j}-2V(q)(dq^{0})^{2}+2dq^{0}dq^{N+1} \ . \label{11_mio}
\end{equation} 
This is the Eisenhart metric \cite{Eisenhart}: its metric tensor will be referred to as
$g_E$ and its components are 
\beq
g_E = \left( 
\begin{array}{ccccc} 
-2 V(q)& 0       & \cdots        & 0     & 1     \\ 
0       & 1& \cdots        & 0& 0     \\ 
\vdots  & \vdots& \ddots        & \vdots& \vdots\\ 
0       & 0& \cdots        & 1& 0     \\ 
1       & 0     & \cdots        & 0     & 0     \\ 
\end{array} \right) 
\label{g_E}
\eeq
as can be derived by Eq.\ (\ref{11_mio}).

The connection between the geodesics of this metric and the natural motions of the
system is contained in a theorem (for a detailed statement and proof see \cite{Lich})
stating that the natural motions of a Hamiltonian dynamical system are obtained by
projecting on the  configuration space-time $M\times\mathbb{R}$ those geodesics of
$(M\times\mathbb{R}^{2},g_{E})$ whose
arc-lengths are positive definite and affinely parametrized with time (remember that $q^0
= t$), i.e., given by:
\begin{equation}
ds^{2}=c_1^{2}\,dt^{2} \ , \label{12_mio}
\end{equation}
where $c_1$ is an arbitrary constant. Condition (\ref{12_mio}) can be equivalently cast as
a condition on $q^{N+1}$, that is
\beq
q^{N+1} = \frac{c_1^2}{2} t + c_2^2 - \int_0^t L\, dt~, \label{13_mio}
\eeq  
where $c_2$ is another arbitrary constant. Conversely, every point of 
$M\times\mathbb{R}^{2}$ such that its projection on the configuration spacetime 
lies on a a physical trajectory of the system and $q^{N+1}$ is given by (\ref{13_mio})
belongs to an affinely parametrized  geodesic of $(M\times\mathbb{R}^{2},g_{E})$.

The nonzero Christoffel symbols of the Eisenhart metric are
\begin{equation}
\Gamma_{00}^{i}=-\Gamma_{0i}^{N+1}=\frac{\partial V}{\partial q_i} ~, \label{13_mio_bis}
\end{equation} 
so that the geodesic equations (\ref{9_mio}) become
\begin{eqnarray} 
\frac{d^2{q^{0}}}{ds^{2}}&=&0 \ , \label{14_mio} \\ 
\frac{d^{2}q^{i}}{ds^{2}}+\Gamma_{00}^{i}\frac{dq^{0}}{ds}\frac{dq^{0}}{ds}&=&0 \ , \label{15_mio} \\ 
\frac{d^{2}q^{N+1}}{ds^{2}}+\Gamma_{0i}^{N+1}\frac{dq^{0}}{ds}\frac{dq^{i}}{ds}&=&0 \ , \label{16_mio} \\  \nonumber
\end{eqnarray} 
and using $ds=dt$, i.e., condition (\ref{12_mio}) with $c_1=1$, we have
\begin{eqnarray} 
\frac{d^{2}q^{0}}{dt^{2}}&=&0 \ \label{17_mio} ; \\ 
\frac{d^{2}q^{i}}{dt^{2}}&=&-\frac{\partial V}{\partial q_{i}} \ \label{18_mio} ; \\ 
\frac{d^{2}q^{N+1}}{dt^{2}}&=&-\frac{dL}{dt} \ . \label{19_mio} \\ \nonumber
\end{eqnarray} 
Equation (\ref{17_mio}) states that $q^{0}=t$, 
the $N$ equations (\ref{18_mio}) are Newton's equations and
Eq.\ (\ref{19_mio}) is the differential version of Equation (\ref{13_mio}). 

The nonvanishing components of the curvature tensor are
\begin{equation}
R_{0i0j}=\partial_{i}\partial_{j}V~; \label{20_mio}
\end{equation} 
it can then be shown that the Ricci curvature (\ref{riccicurv}) in the direction of
motion, i.e., in the direction of the velocity vector $v$ of the geodesic, is given by
\begin{equation}
K_{R}(v)=\triangle V \label{22_mio}~,
\end{equation} 
where $\triangle V$ is the Laplacian of the potential $V$, 
and that the scalar curvature ${\cal R}$ identically vanishes. 

We note that $K_{R}(v)$ is nothing but a scalar measure of the average curvature ``felt'' by the
system during its evolution; we will refer to it simply as $K_R$ dropping the dependence
on the direction. Another feature of $K_{R}$ is its very simple analytical
expression which simplifies both analytical calculation and numerical
estimates. It is also worth noticing that expression (\ref{22_mio}) is a very natural and
intuitive measure of the curvature of the energy landscape, as it can be seen as a naive
generalization of the curvature $f''(x)$ of the graph of a one-variable function  to the
graph of the $N$-dimensional function  $V(q_{1},...,q_{N})$: the Laplacian of the
function. However, the previous discussion shows that it is much more than a naive
measure of curvature and that it contains information on the local neighborhood of the
dynamical trajectories. This can be exploited to gain information on the {\em stability}
of the dynamics. However, we shall not go on along this line here, and we refer the
reader to the review \cite{physrep} as well as to the monograph \cite{marco_book}.

The Ricci curvature defined in Eq.\ (\ref{22_mio}) will be the quantity we shall use to
characterize the geometry of the energy landscape.

\section{Model and numerical simulations}

Let us now describe the model whose energy landscape geometry we studied. We considered a
simple model able to describe protein-like polymers as well as polymers with no tendency
to fold; the different behaviors being selected upon the choice of the amino acidic
sequence. The  model we chose is a minimalistic model originally introduced by Thirumalai
and coworkers \cite{Thirumalai}. In order to characterize its energy landscape geometry, 
we sampled the value of the Ricci curvature $K_R$ defined in Eq.\ (\ref{22_mio})   along
its dynamical trajectories. We shall now describe the model and the details of the
numerical simulations; then we shall report on the results.

\subsection{The model}

The Thirumalai model is a three-dimensional off-lattice model of a
polypeptide which has only three different
kinds of amino acids: polar (P), hydrophobic (H) and neutral (N). The potential
energy is 
\bea
V(\vec{r}_1,\ldots,\vec{r}_N) & = & V_\text{bond}(|{\vec{r}}_i - {\vec{r}}_{i-1}|) +
V_\text{angular}(|{\vartheta}_i - {\vartheta}_{i-1}|) \nonumber \\
& +& 
V_\text{dihedral}({\psi}_i) + V_\text{non-bonded}({\vec{r}}_1,\ldots,{\vec{r}}_N)
\eea 
where 
\bea
V_\text{bond} & = & \sum_{i=1}^{N-1} \frac{k_r}{2} (|{\vec{r}}_i - {\vec{r}}_{i-1}| 
- a)^2 ~; \label{bond}\\
V_\text{angular} & = & \sum_{i=1}^{N-2} \frac{k_\vartheta}{2} (|{\vartheta}_i -
{\vartheta}_{i-1}| - \vartheta_0)^2~; \label{ang}\\
V_\text{dihedral} & = & \sum_{i=1}^{N-3} \left\{A_i [1 + \cos \psi_i] + B_i[1 +
\cos(3\psi_i)]  \right\}~; \\
V_\text{non-bonded} & = & \sum_{i= 1}^{N-3} \sum_{j= i+3}^{N}
V_{ij}(|{\vec{r}}_{i,j}|)~, \label{non-bonded}
\eea
where $\vec{r}_i$ is the position vector of the $i$-th monomer, 
$\vec{r}_{i,j} = \vec{r}_i - \vec{r}_j$, $\vartheta_i$ 
is the $i$-th bond angle, i.e., the 
angle between $\vec{r}_{i+1}$ and $\vec{r}_{i}$, $\psi_i$ the $i$-th dihedral
angle, that is the angle between the vectors $\hat{n}_i = \vec{r}_{i+1,i} \times
\vec{r}_{i+1,i+2}$ and  $\hat{n}_{i+1} = \vec{r}_{i+2,i+1} \times
\vec{r}_{i+2,i+3}$, $k_r = 100$, $a = 1$, $k_\vartheta = 20$, $\vartheta_0 =
105^\circ$, $A_i=0$ and $B_i = 0.2$ if at least two among the residues
$i,i+1,i+2,i+3$ are N, $A_i = B_i = 1.2$ otherwise. As to $V_{ij}$, we have
\beq 
V_{ij} = \frac{8}{3}\left[\left(\frac{a}{r} \right)^{12} + \left(\frac{a}{r}
\right)^{6} \right]
\eeq
if $i,j = \text{P},\text{P}$ or $i,j = \text{P},\text{H}$,  
\beq
V_{ij} = 4\left[\left(\frac{a}{r} \right)^{12} - \left(\frac{a}{r}
\right)^{6} \right]
\eeq
if $i,j = \text{H},\text{H}$ and  
\beq
V_{ij} = 4\left(\frac{a}{r}
\right)^{6}
\eeq if either $i$ or $j$ are N \cite{Thirumalai}. 
\begin{figure}
\begin{center}
\psfrag{ylabel}{$V_{\text{dihedral}} (\psi)$}
\psfrag{xlabel}{$\psi ~(\text{rad})$}
\includegraphics[width=8cm,angle=-90,clip=true]{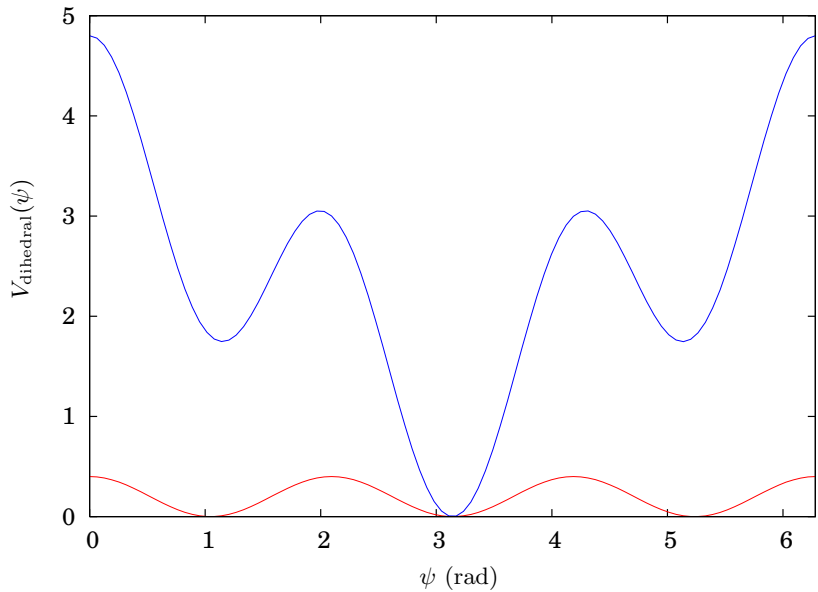}
\end{center}
\caption{(Color online) V$_\text{dihedral}$ vs.\ dihedral angle $\psi_i$: the
red (lower) curve corresponds to
the case in which at least two out of the four beads defining the angle $\psi_i$ are
neutral (N), the
blue (upper) curve to the other cases; the central minimum corresponds to a planar
configuration of the four monomers defining $\psi_i$.}
\label{figure_die}
\end{figure}

The meaning of the different terms of the potential energy is the following.
V$_\text{bond}$ accounts for the covalent bonds between the $\alpha$-carbons on the main
protein chain, V$_\text{angular}$ for the energy cost associated to deviations from the
preferred bond angle between two neighbors on the chain, and V$_\text{dihedral}$
describes the contribution of non planar structures to the energy. These terms have very
different energy scales, due to the different strengths of the bonds, and the dihedral
term V$_\text{dihedral}$ strongly depends on the local amino acidic composition: the
formation of non-planar structures (turns and so on) is favored by the presence of
neutral monomers (see Fig.\ \ref{figure_die}). All the interactions between
non-neighboring monomers along the chain, including hydrophobic effective interactions,
are described by the Lennard-Jones term $V_\text{non-bonded}$. The presence of different
energy scales as well as the competition between bonding terms (favoring planar, straight
configurations) and non-bonded interactions (favoring compact configurations when the
polymer is mostly hydrophobic) suggest that the energy landscape is very complicated and
the possibilty of a high degree of frustration in the system: however, it is clear that
these features will strongly depend on the details of the aminoacidic sequence. For
example, a hydrophobic homopolymer is expected to be more frustrated than a 
heteropolymer, because in the homopolymer case there is a large number of compact
configurations which are energetically equivalent (at least as long as the non-bonded
contribution to the energy is considered) while with a heteropolymer different compact
configurations have different energies.

\subsection{Simulations and results}

Although the identity between trajectories  and projections of the geodesics of $(M,g_E)$
only holds if the dynamics is the Newtonian one,  a Langevin dynamics, obtained by adding
to the deterministic force $\nabla V$ a  random force according to the
fluctuation-dissipation theorem and a friction term proportional  to the velocity, is a
more reasonable model of the dynamics of a polymer in aqueous solution when the solvent
degrees of freedom are not taken into account explicitly. Since we are interested not in
the details of the time series of $K_R$ along a particular trajectory but only in its
statistical distribution, we may expect that also a sampling obtained using the Langevin
dynamics gives the same information on the geometry of the landscape.  To check this
assumption  we let the system evolve with both a Newtonian dynamics (using a symplectic
algorithm \cite{MacLachlan} to integrate the equations of motion) and a Langevin dynamics
(using the same algorithm---a modified  Verlet---and parameters as 
in Ref.\ \cite{Thirumalai}) 
obtaining very similar results. This is reasonable because at
equilibrium, i.e., for sufficiently long simulations, the Langevin dynamics 
(resp.\ Newtonian dynamics) samples the
phase space according to a canonical (resp.\ microcanonical) distribution, and the two
distributions are expected to be equivalent for large systems\footnote{The interactions
are all well-behaved and sufficiently short-range as to guarantee equivalence of statistical ensembles in
the thermodynamic limit.}, apart from small deviations due to finite size effects. 

In the following we shall refer only to results obtained with
Langevin dynamics. 

\subsubsection{Simulation details}

All the numerical results will be given in natural units defined as follows: the unit of length
$\ell$ is equal to the equilibrium bond length (i.e., the equilibrium distance between two
consecutive beads in the chain), the unit of energy $\varepsilon$ is the hydrophobic interaction
scale, i.e.,  the depth of the Lennard-Jones potential well between two hydrophobic beads, the
unit of mass $m$ is the mass of the residues. Then the time unit becomes $\ell
\sqrt{m/\varepsilon}$ and the temperature unit is $\varepsilon/k_B$ where $k_B$ is the Boltzmann
constant. 

As already mentioned above, the integration algorithm we used to solve
numerically the Langevin equations of motion is the
modified Verlet given in Ref.\ \cite{Thirumalai}. The time step was $\Delta t = 5\times
10^{-3}$. Each run was $1.7 \times 10^7$ time steps long, including
equilibration. Results for a given temperature were obtained averaging over 8 different
randomly chosen initial conditions. The mean and rms fluctuation of the observables, and in
particular of the curvature of the landscape $K_R$, was estimated from the 
histogram of the sampled values. Statistical errors on the single run have been estimated
dividing each run in 12 tranches, considered as independent, and then propagated to the
final result.

\subsubsection{Sequences}

We considered six different sequences of ``amino acids'' H, P and N: four of  22 monomers,
$S^{22}_{\text{g}}$,  $S^{22}_{\text{b}}$,  $S^{22}_{\text{i}}$,  $S^{22}_{\text{h}}$, 
and two sequences of 46 monomers, $S^{46}_{\text{h}}$, $S^{46}_{\text{g}}$. The six
sequences are listed in Table \ref{table}. Sequences $S^{22}_{\text{g}}$ and
$S^{46}_{\text{g}}$ had already  been identified as good folders \cite{Thirumalai,
Thirumalai2} and our simulations confirmed this finding: below a given temperature
$S^{22}_{\text{g}}$ (resp.\ $S^{46}_{\text{g}}$) always reached the same 
$\beta$-sheet-like structure (resp.\ $\beta$-barrel-like structure). A plot of the number
of native contacts $N_n$ (see Sec.\ \ref{tfold} for the definition) 
as a function of the temperature for these two proteinlike
sequences is reported in Fig.\ \ref{figure_cont_nat}. 
Homopolymers
$S^{22}_{\text{h}}$ and $S^{46}_{\text{h}}$, on the other hand,  showed a hydrophobic
collapse but no tendency to reach a particular configuration in the collapsed phase, as
expected.
Sequence $S^{22}_{\text{b}}$ (which has the same overall composition of
$S^{22}_{\text{g}}$ rearranged in a different sequence) behaved as a bad folder and did
not reach  a unique native state, while $S^{22}_{\text{i}}$ was constructed by us to show
a somehow intermediate behavior between good and bad folders: it always forms the same
structure involving the middle of the sequence, while the beginning and the end of the
chain fluctuate also at low temperature.

\begin{table}
\begin{tabular}{|c|cl|}
\hline
name & & sequence \\
\hline
S$^{22}_{\text{g}}$ & & PH$_9$(NP)$_2$NHPH$_3$PH \\
S$^{22}_{\text{b}}$ & & PHNPH$_3$NHNH$_4$(PH$_2$)$_2$PH \\ 
S$^{22}_{\text{i}}$ & & P$_4$H$_5$NHN$_2$H$_6$P$_3$ \\ 
S$^{22}_{\text{h}}$ & & H$_{22}$ \\
S$^{46}_{\text{g}}$ & & 
P(HP)$_5$N$_{3}$H$_{9}$N$_{3}$(HP)$_4$N$_{3}$H$_{9}$ \\
S$^{46}_{\text{h}}$ & & H$_{46}$ \\
\hline
\end{tabular}
\caption{The six sequences considered. (X)$_y$ means that X is repeated $y$
times.}
\label{table}
\end{table}

\subsubsection{Estimate of collapse and folding temperatures}

\label{tfold}

The hydrophobic collapse temperature $T_\theta$ and the folding temperature $T_f$
(the latter only for the two protein-like sequences S$_\text{g}^{22}$ and 
S$_\text{g}^{46}$ and for the ``intermediate'' sequence
S$_\text{i}^{22}$) were estimated in a standard way as follows. 

$T_\theta$ has been estimated as the temperature
where the specifc heat shows a peak. The rationale for this definition is that the
hydrophobic collapse of a polymer becomes a thermodynamic phase transition, commonly
referred to as $\theta$ transition, in the
thermodynamic limit of an infinite number of monomers \cite{degennes}; at the critical
temperature (usually denoted by $T_\theta$, notation that we adopt also for our finite
size case) the specific heat diverges, and the specific heat peak we observed in our
systems is nothing but the precursor of this divergence. One could define a collapse
temperature also by monitoring quantities directly related to the collapse itself, as the
gyration radius or the end-to-end distance of the polymer; we checked that the estimates
of $T_\theta$ obtained this way were perfectly compatible with those based on the
specific heat.

\begin{figure}[b]
\begin{center}
\includegraphics[width=5.75cm,angle=-90,clip=true]{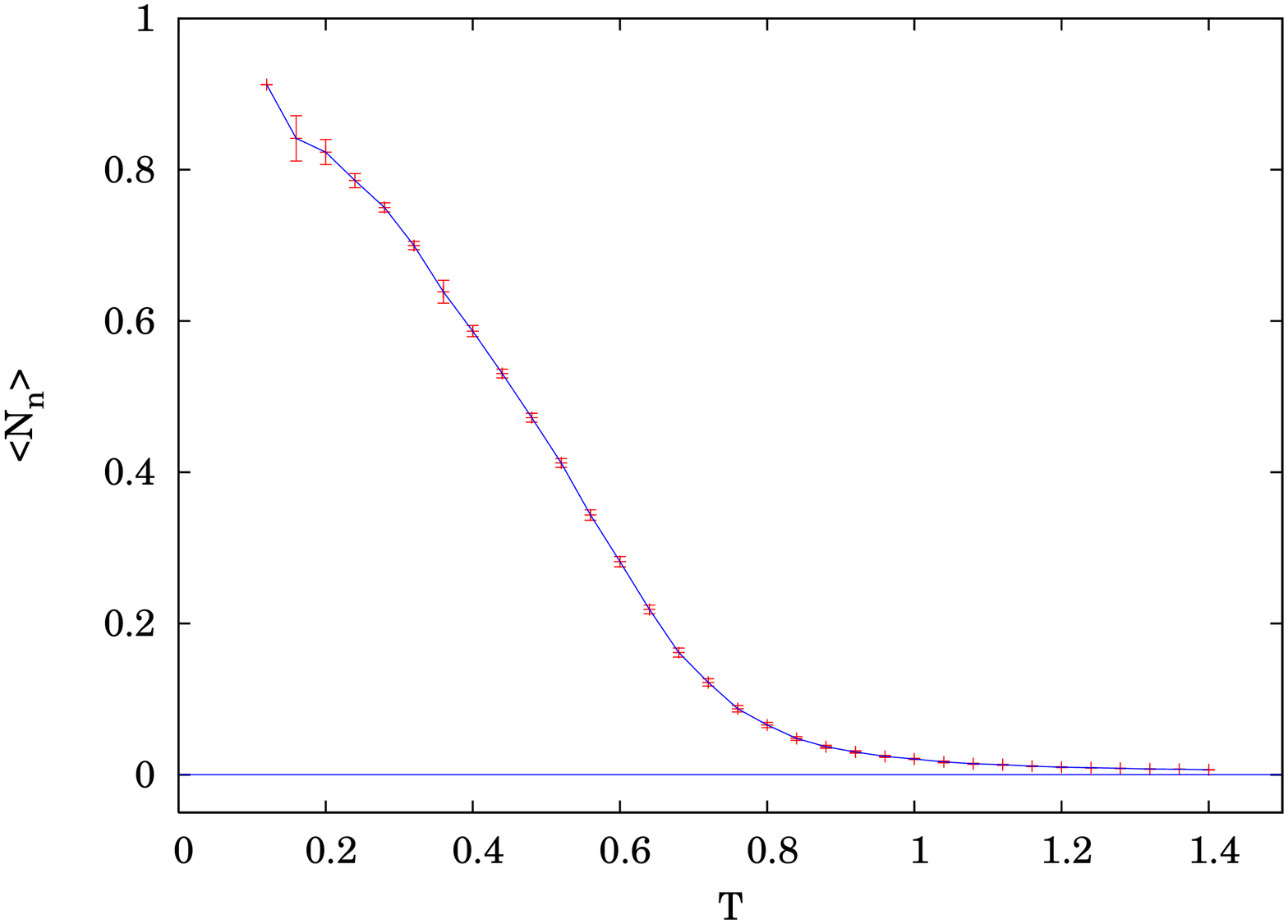}
\includegraphics[width=5.75cm,angle=-90,clip=true]{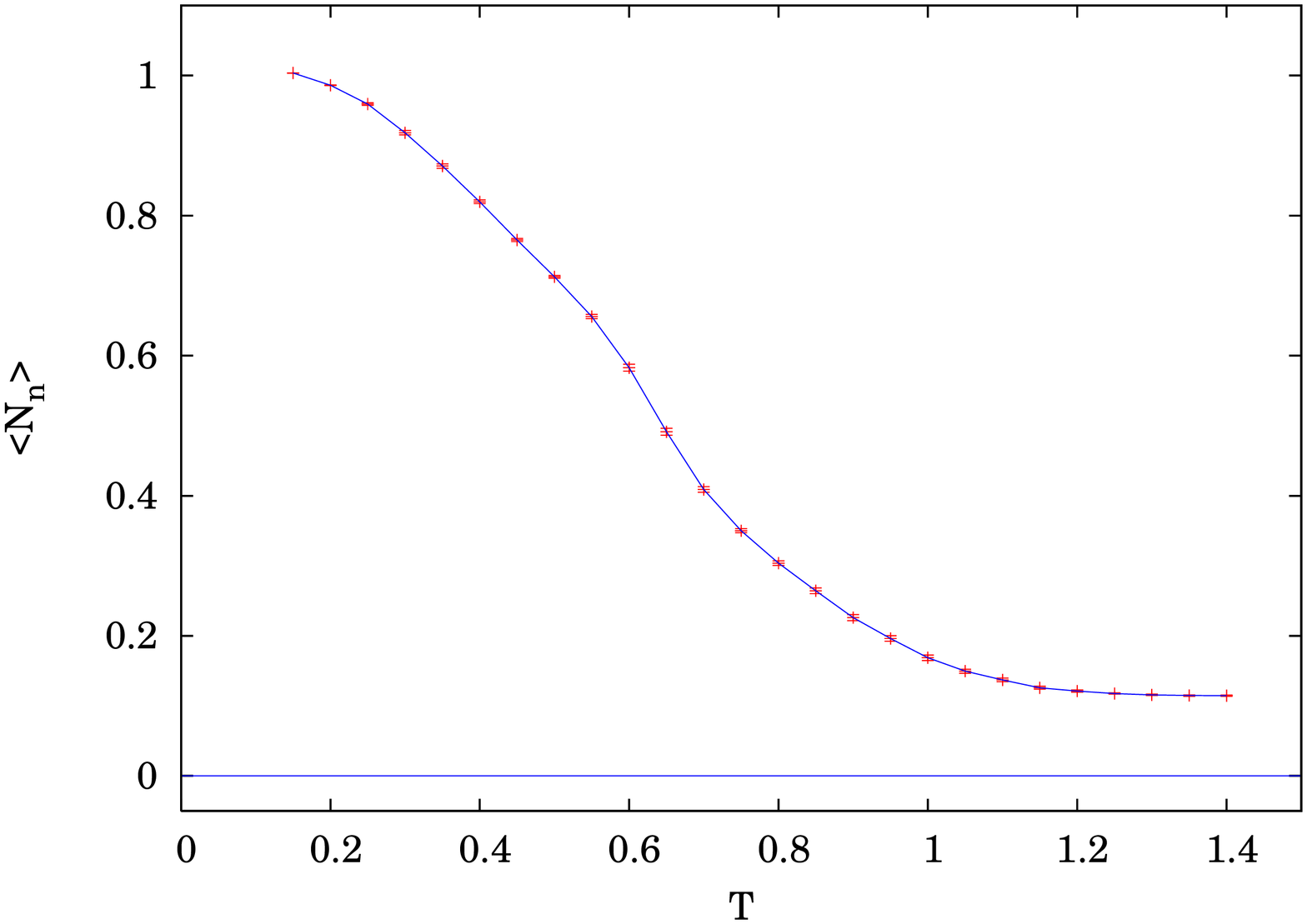}
\end{center}
\caption{(Color online) $\langle N_{n} \rangle$ vs.\ temperature $T$ for the
two good folders: $S^{22}_{\text{g}}$ (left), $S^{46}_{\text{g}}$ (right). The curves are a guide to the eye.}
\label{figure_cont_nat}
\end{figure}

At variance with the collapse transition, the folding transition does not have any
corresponding ``true'' phase transition in the thermodynamic limit, simply
because no thermodynamic limit exists for a protein: we
cannot increase the number of amino acids of a protein beyond any limit 
whithout destroying its protein-like behavior \cite{Dill,CecconiBurioni}. From the point
of view of statistical physics, 
protein folding may be considered a genuine finite-size phenomenon. Any estimate of the
folding temperature $T_f$ will then be somehow fuzzy. We adopted one of the various
possible protocols. First, we defined the fraction $N_n$ 
of native contacts in any configuration as
\begin{equation}
N_{n}=\frac{1}{n_{c}}\sum_{i=1}^{N}\sum_{j=i+2}^{N}\Theta(d_{s}-r_{ij}) 
\Theta(d_{s}-r^n_{ij})~, 
\label{mio_ij}
\end{equation} 
where $r_{ij}$ is the distance between the $i$-th and $j$-th residues, 
$r_{ij}^{n}$ the distance between the same residues in the reference (native) 
configuration, $n_{c} = \sum_{i=1}^{N}\sum_{j=i+2}^{N}\Theta(d_{s}-r^n_{ij})$ 
the total number of
contacts in the native configuration, and $d_{s}$ the threshold distance below which two
beads are considered in contact: the value of $d_{s}$ is the mean distance between beads
in contact in the native configuration. Then, $T_{f}$ is defined as 
the temperature of the inflection point in the curve $N_{n}(T)$. These curves are shown
in Fig.\ \ref{figure_cont_nat}. 
The estimated $T_f$ and $T_\theta$ for the six sequences are given in Table
\ref{table_temp}. Errors on $T_\theta$ (resp.\ $T_f$) have been obtained by estimating the
interval of the position of the peak in $c_V(T)$ (resp.\ in the derivative of $N_n(T)$)
compatible with the errors on the numerical data for $c_V$ (resp.\ $N_n$).  
\begin{table}
\begin{tabular}{|c|c|c|}
\hline
sequence & $T_\theta$ & $T_f$ \\
\hline
S$^{22}_{\text{g}}$ & $0.65 \pm 0.05$ & $0.55 \pm 0.1$ \\
S$^{22}_{\text{b}}$ & $0.55 \pm 0.15$ &  none \\ 
S$^{22}_{\text{i}}$ & $0.75 \pm 0.1$ & $0.7 \pm 0.2$\protect\footnote{For this sequence a ``quasi-folding''
transition where only half of the sequence folds is detected (see text).} \\ 
S$^{22}_{\text{h}}$ & $0.65 \pm 0.05$ & none \\
S$^{46}_{\text{g}}$ & $0.65 \pm 0.025$ & $0.65 \pm 0.05$ \\
S$^{46}_{\text{h}}$ & $0.70 \pm 0.025$ &  none \\
\hline
\end{tabular}
\caption{The collapse ($T_\theta$) and folding ($T_f$) temperatures for the 
six sequences considered.}
\label{table_temp}
\end{table}
We note that, for the two sequences where folding occurs, $T_f \approx T_\theta$ (indeed
$T_f = T_\theta$ within the estimated errors). This is
not surprising at all since in general $T_f \leq T_\theta$, and one expects the folding
to be more efficient if $T_f/T_\theta \approx 1$ because in this situation the polymer
approaches the native state immediately when the temperature is lowered below $T_\theta$,
reducing the possibility of misfolding in non-native compact configurations \cite{keyes}.  

\subsubsection{Thermodynamic and geometric observables}

\begin{figure}[h]
\begin{center}
\includegraphics[width=8cm,clip=true]{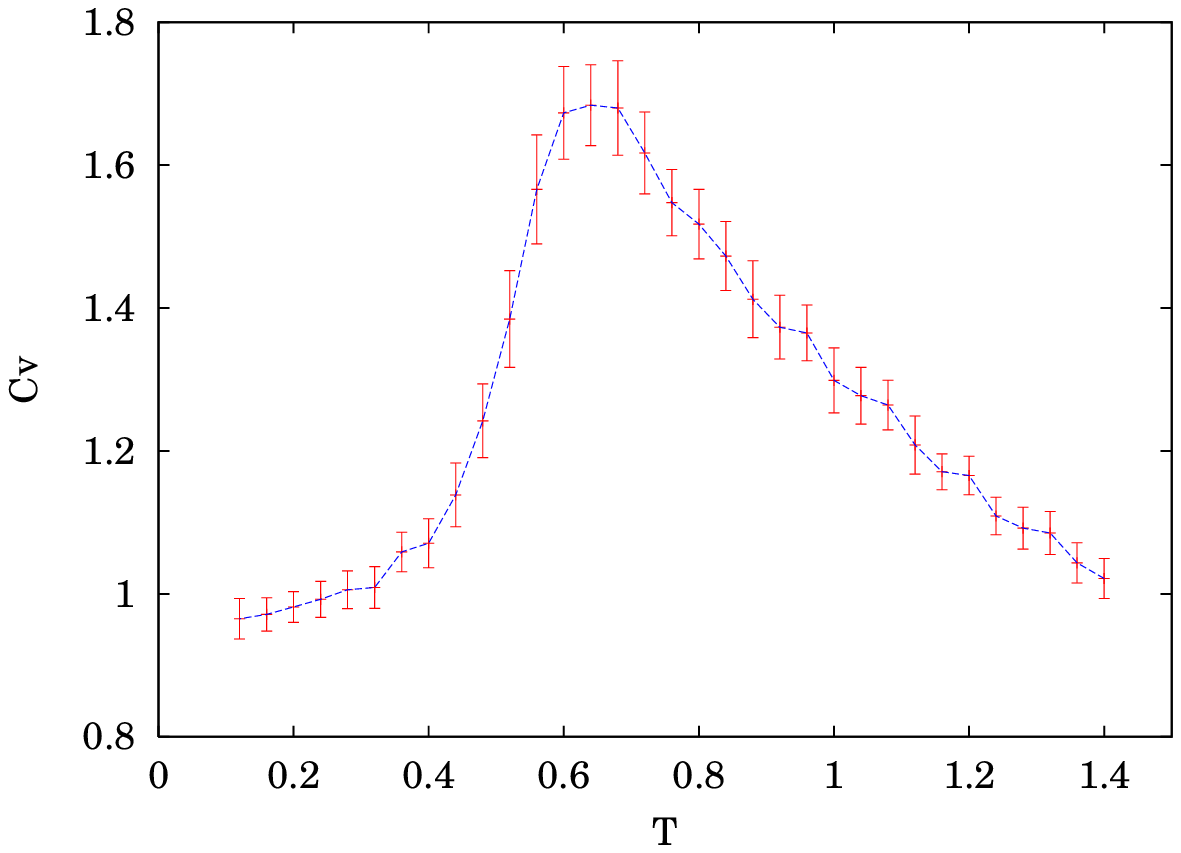}
\includegraphics[width=8cm,clip=true]{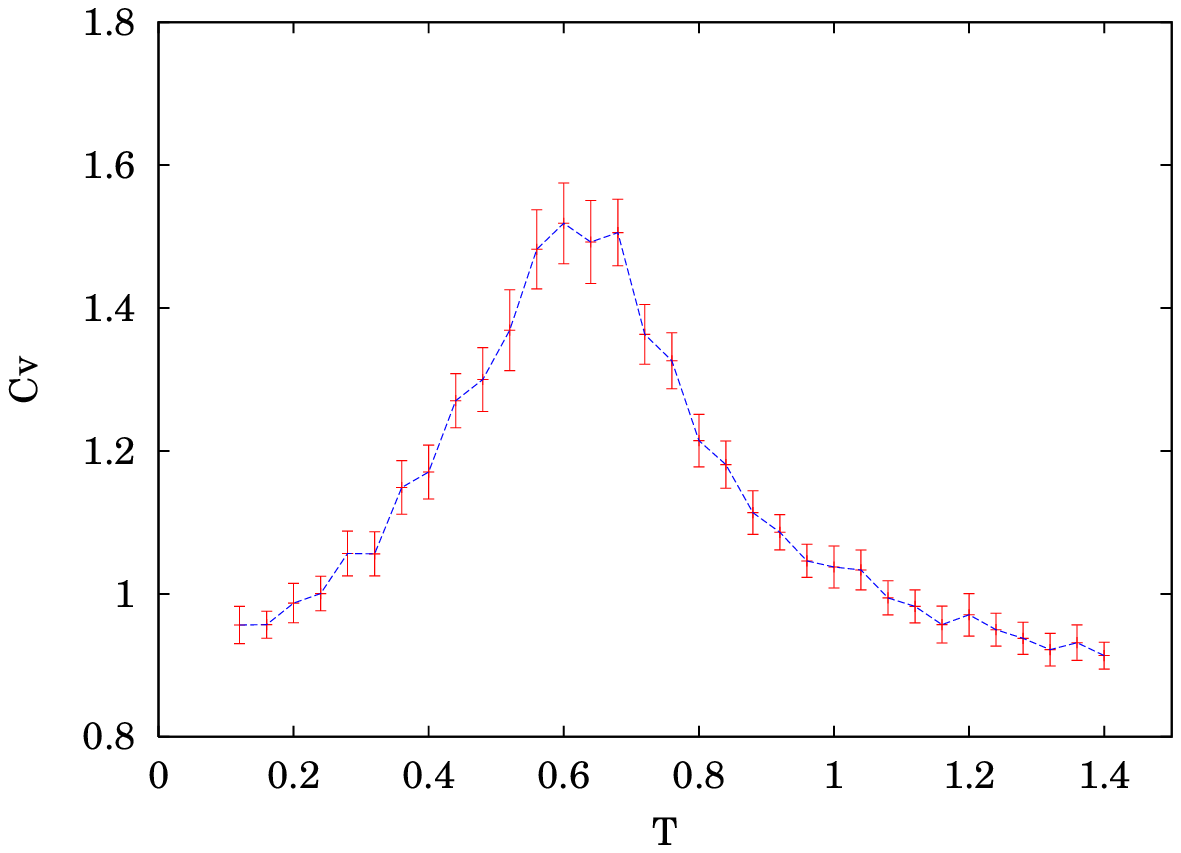}
\end{center}
\caption{(Color online) Specific heat $c_V$ vs.\ temperature $T$ for the
homopolymer $S^{22}_{\text{h}}$ (left) and for the good folder
$S^{22}_{\text{g}}$ (right). The curves are a guide to the eye.}
\label{figure_cv}
\end{figure}

\begin{figure}[h]
\begin{center}
\includegraphics[width=5.75cm,angle=-90,clip=true]{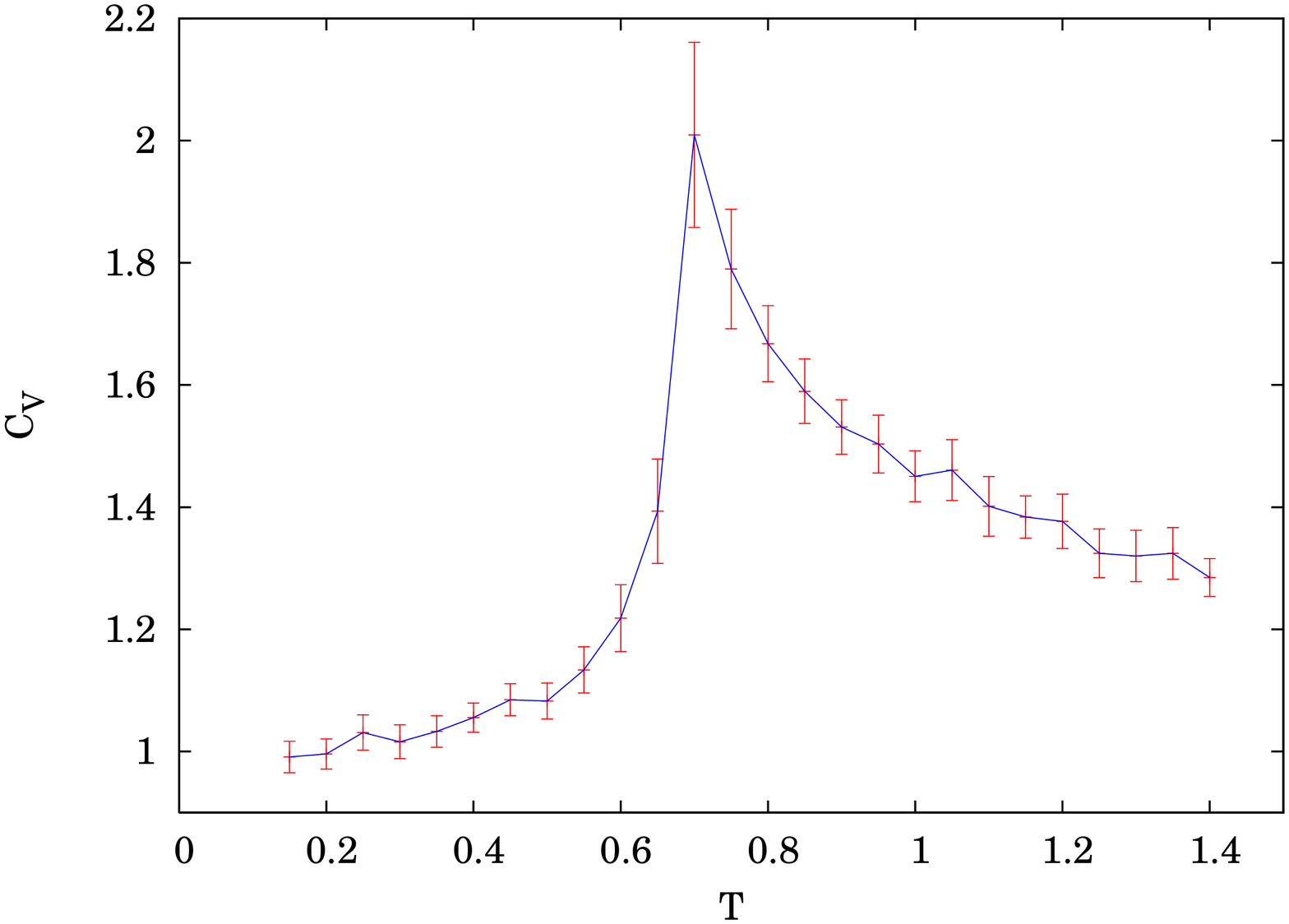}
\includegraphics[width=5.75cm,angle=-90,clip=true]{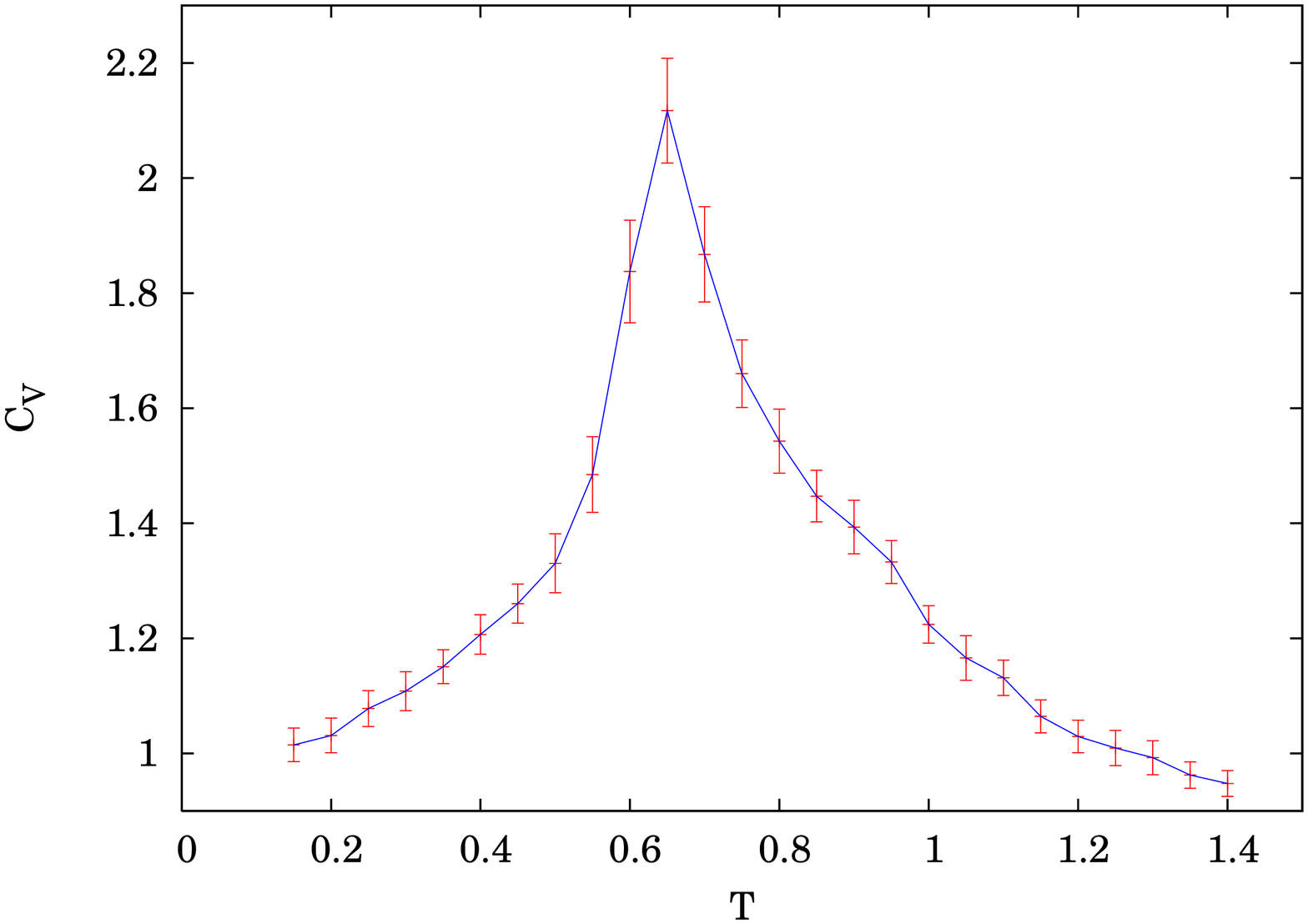}
\end{center}
\caption{(Color online) As in Fig.\ \protect\ref{figure_cv} for the homopolymer $S^{46}_{\text{h}}$ (left) and 
the good folder $S^{46}_{\text{g}}$ (right).}
\label{figure_cv_long}
\end{figure}

\begin{figure}[h]
\begin{center}
\includegraphics[width=8cm,clip=true]{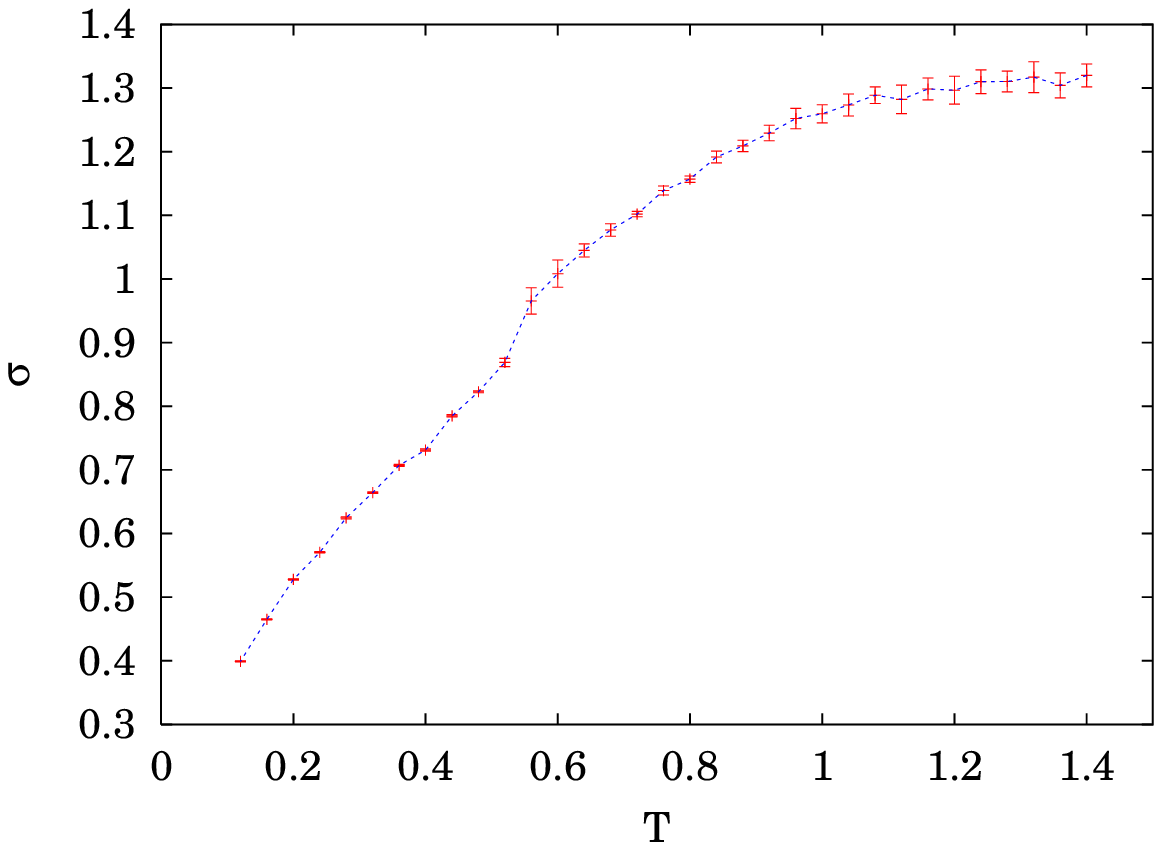}
\includegraphics[width=8cm,clip=true]{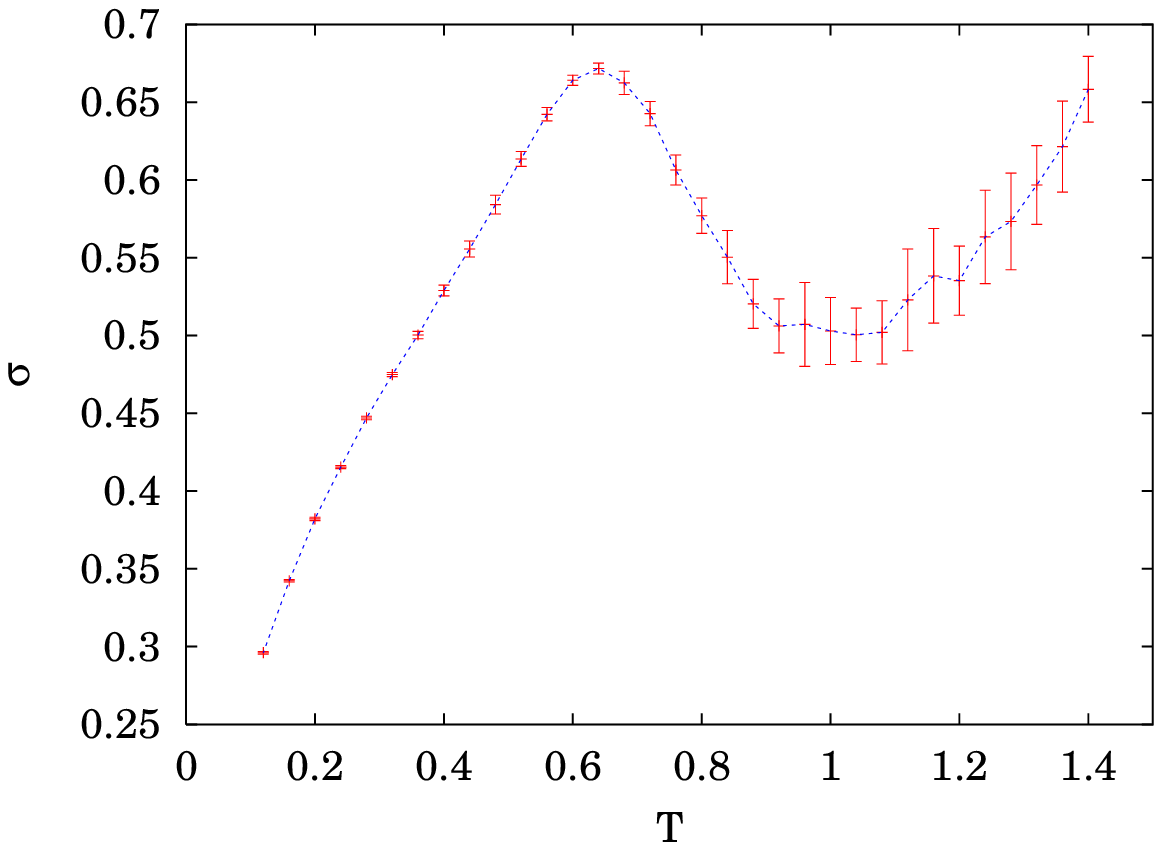}
\end{center}
\caption{(Color online) Relative curvature fluctuation $\sigma$ vs.\ temperature $T$ for the
homopolymer $S^{22}_{\text{h}}$ (left) and for the good folder
$S^{22}_{\text{g}}$ (right). The curves are a guide to the eye.}
\label{figure_sigma}
\end{figure}

\begin{figure}
\begin{center}
\includegraphics[width=5.75cm,angle=-90,clip=true]{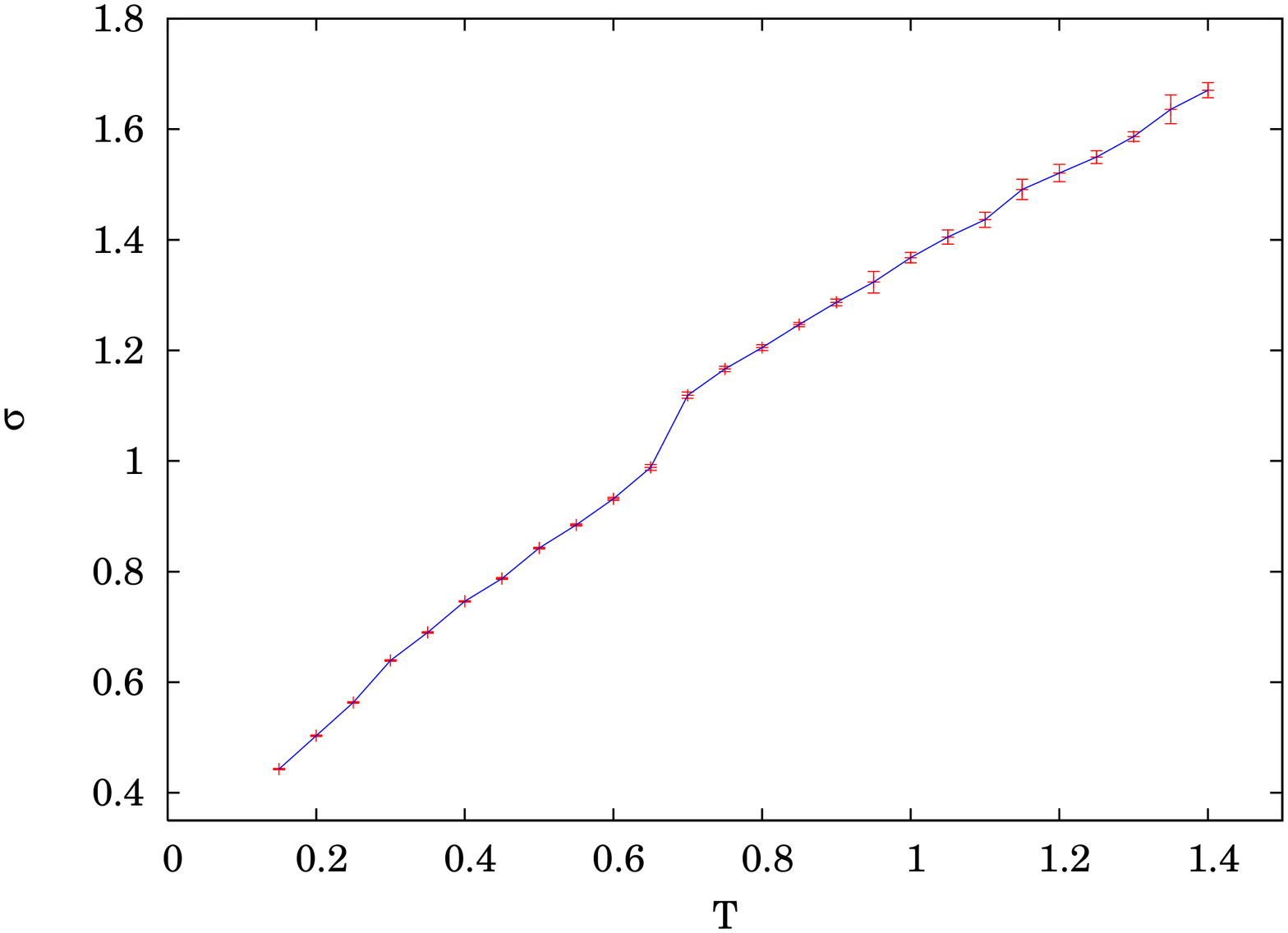}
\includegraphics[width=5.75cm,angle=-90,clip=true]{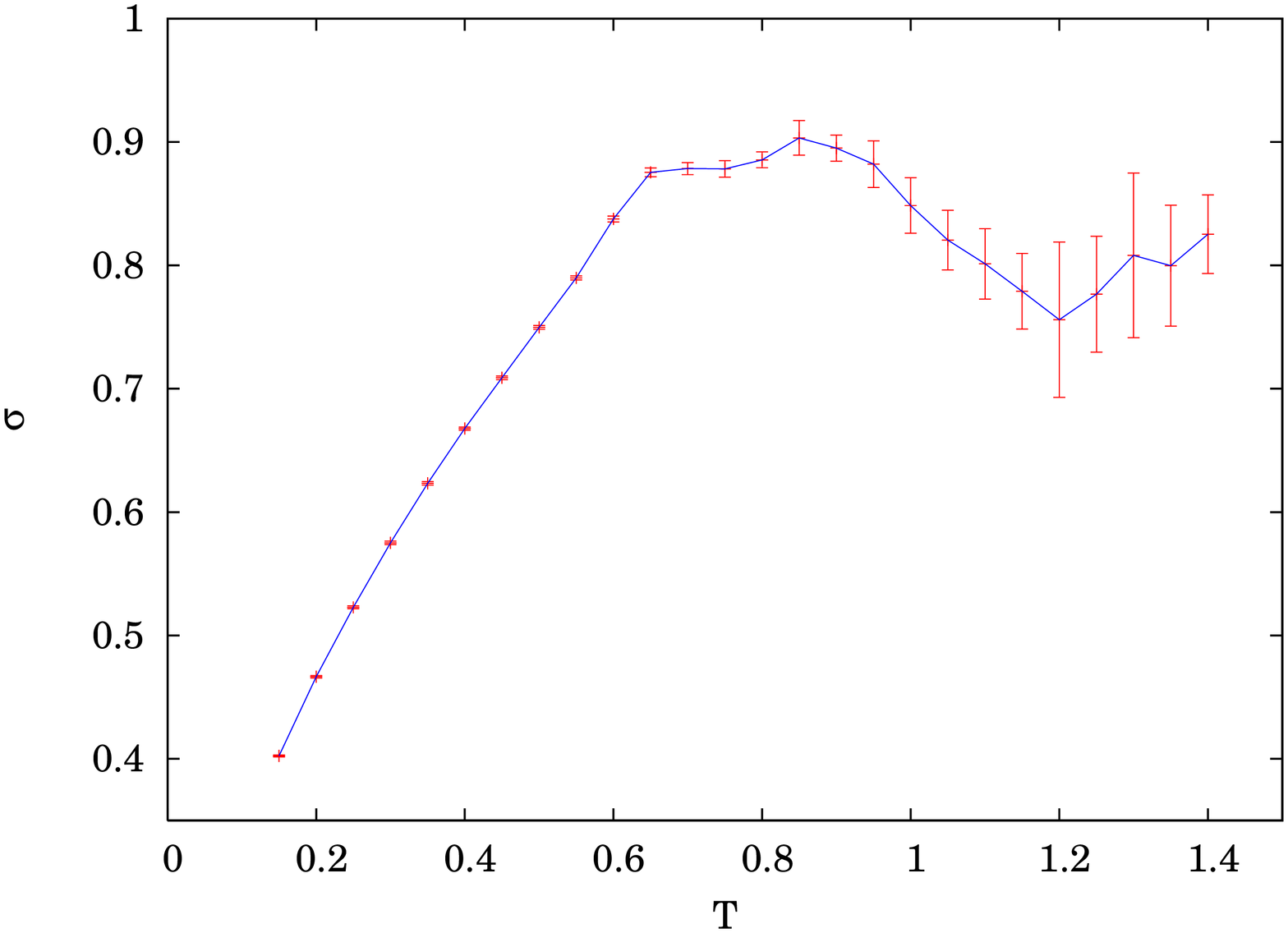}
\end{center}
\caption{(Color online) As in Fig.\ \protect\ref{figure_sigma} for the homopolymer $S^{46}_{\text{h}}$
and the proteinlike sequence $S^{46}_{\text{g}}$.}
\label{figure_sigma_long}
\end{figure}

As to standard thermodynamic observables, all the sequences showed very similar
behaviors: to give an example, 
in Fig.\ \ref{figure_cv} we compare the specific heat $c_V$ of the
homopolymer $S^{22}_{\text{h}}$ and of the good folder $S^{22}_{\text{g}}$: both
exhibit a peak at the transition, and on the sole 
basis of this picture it would be
hard to discriminate between a simple hydrophobic collapse and a folding. The same
happens in the case of the longer homopolymer $S^{46}_{\text{h}}$ and the good folder
$S^{46}_{\text{g}}$  (Fig.\ \ref{figure_cv_long}).
On the other hand, a dramatic difference between the homopolymers and the good
folders shows up if we consider the geometric properties of the landscape. In
particular, a lot of information appears to be encoded in the fluctuations of the
Ricci curvature $K_R$, i.e., of the Laplacian of
the potential energy---see Eq.\ (\ref{22_mio}). 
We defined a relative adimensional curvature fluctuation
$\sigma$ as
\beq
\sigma = \frac{\sqrt{\frac{1}{N}\left( \langle K_R^2\rangle_t - \langle
K_R\rangle_t^2\right)}}{\frac{1}{N}\langle K_R\rangle_t}
\eeq
where $\langle \cdot \rangle_t$ stands for a time average: 
in Fig.\ \ref{figure_sigma} we plot $\sigma$ as a function of the 
temperature $T$
for the homopolymer $S^{22}_{\text{h}}$ and for the good folder 
$S^{22}_{\text{g}}$. A clear peak shows up in the case of the good folder, close
to the folding temperature $T_f$ 
below which the system is mostly in the native state, while
no particular mark of the hydrophobic collapse can be seen in the case of the
homopolymer. A similar situation happens for the longer sequences, the good folder
$S^{46}_{\text{g}}$ and the homopolymer $S^{46}_{\text{h}}$ 
(Fig.\ \ref{figure_sigma_long}). 


The relative curvature fluctuation $\sigma$ of the energy landscape appears then to be a
good marker of the presence of a folding transition, at least in the simple model 
considered here. We stress that the comparison between sequences of length 22 and 46
clearly indicates that the peak in $\sigma$ is really related to the folding and not to the
hydrophobic collapse, because $\sigma$ for the
long homopolymer $S^{46}_{\text{h}}$ is even smoother than in the case of the short
homopolymer $S^{22}_{\text{h}}$, at variance with the specific heat which
develops a sharper peak, consistently
with the fact that the system is closer to the situation where 
a thermodynamic $\theta$-transition exists.

In the case of the longer sequence (Fig.\ \ref{figure_sigma_long}) 
the peak in the curvature
fluctuations appears more structured: it seems to be the superposition of two peaks, one
at $T_f \approx T_\theta$ and another, even higher, at $T \simeq 0.85$ where the specific
heat shows a ``shoulder''. This may be related to the fact that the good folder 
$S^{46}_{\text{g}}$ seems to reach its native state---a $\beta$-barrel made of two
$\beta$-sheets---in two steps: first, at $T \simeq 0.85$, the two $\beta$-sheets form but
still are free to move the one respect to the other, then, at $T_f$, the two sheets fold
into the native $\beta$-barrel. This interpretation of the folding process as composed of
two steps is corroborated by the results on the gyration radius as a function of $T$
(data not shown). The curvature fluctuations seem then to indicate both
steps of folding with two separate peaks which superimpose on each other to give the structure observed
in Fig.\  \ref{figure_sigma_long}. 

\begin{figure}
\begin{center}
\includegraphics[width=8cm,clip=true]{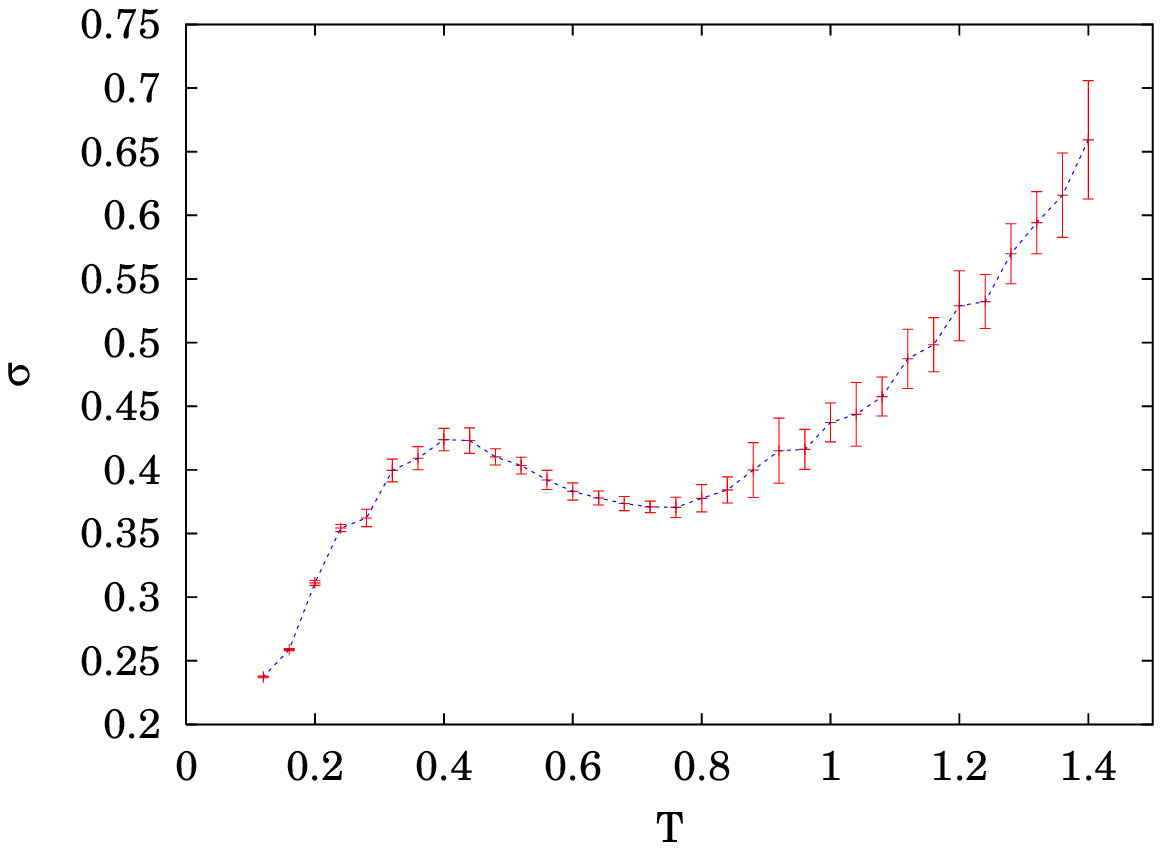}
\includegraphics[width=8cm,clip=true]{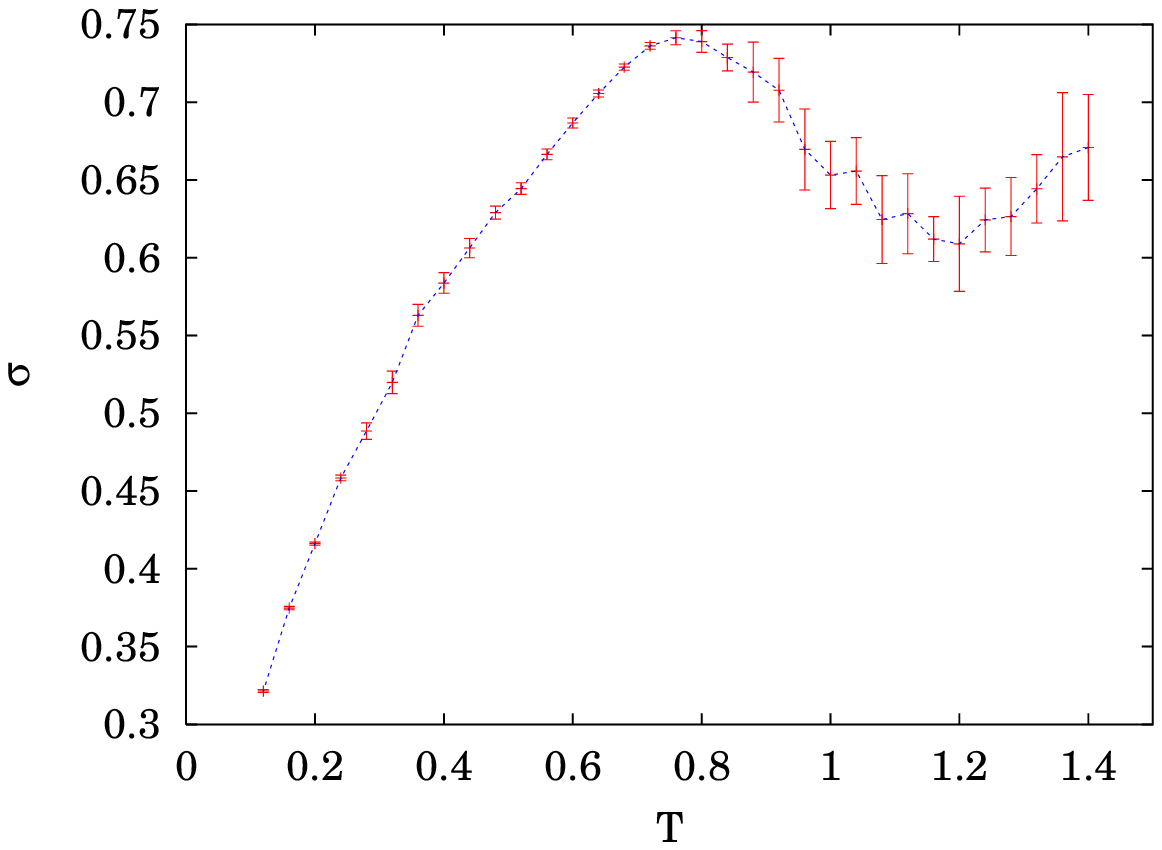}
\end{center}
\caption{(Color online) $\sigma(T)$ for the sequence $S^{22}_{\text{b}}$ (on the left) and for the sequence $S^{22}_{\text{i}}$ (on the right). The curves are a guide to the eye.}
\label{figure_bad}
\end{figure}

As to the other sequences,  
for the bad folder $S^{22}_{\text{b}}$, $\sigma(T)$ is not as
smooth as for the homopolymers, but only a very weak signal is found at $T\simeq 0.4$,
lower than $T_\theta$; below this temperature the systems seems to behave as
a glass\footnote{We could denote it ``glassy temperature'' $T_g$ according to a widespread use,
but we avoid doing that because we did not investigate thorughly the low-temperature
dynamics. Nonetheless on the basis of our findings we expect that in these systems
$T_g\simeq 0.4$.}. For the ``intermediate'' sequence $S^{22}_{\text{i}}$ a peak is present
at the ``quasi-folding'' temperature, although considerably broader than in the
case of $S^{22}_{\text{g}}$ (see Fig.\ \ref{figure_bad}). 

\subsection{A closer look at the curvature of the landscape}

\begin{figure}
\begin{center}
\includegraphics[width=6cm,clip=true,angle=-90]{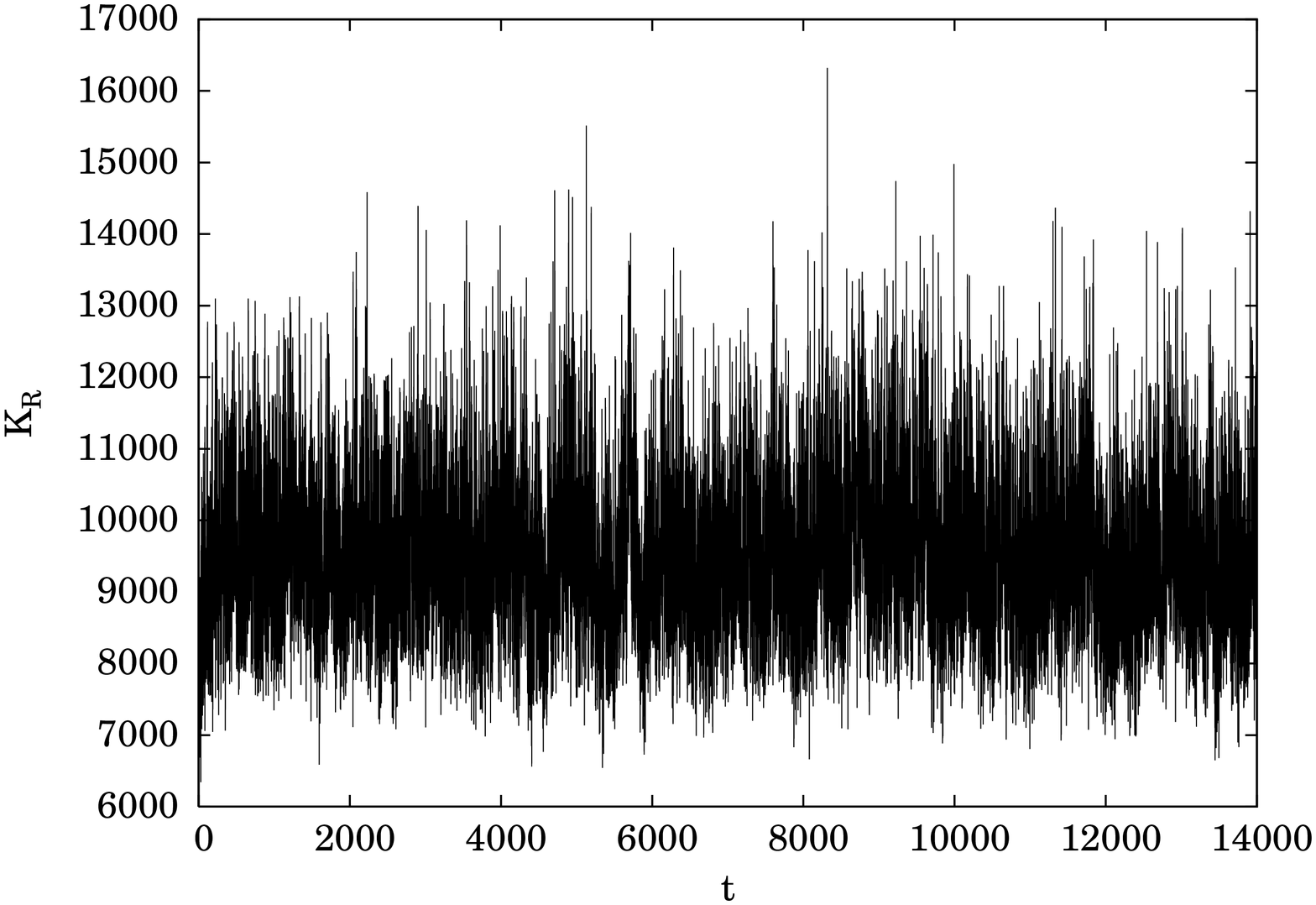}
\includegraphics[width=6cm,clip=true,angle=-90]{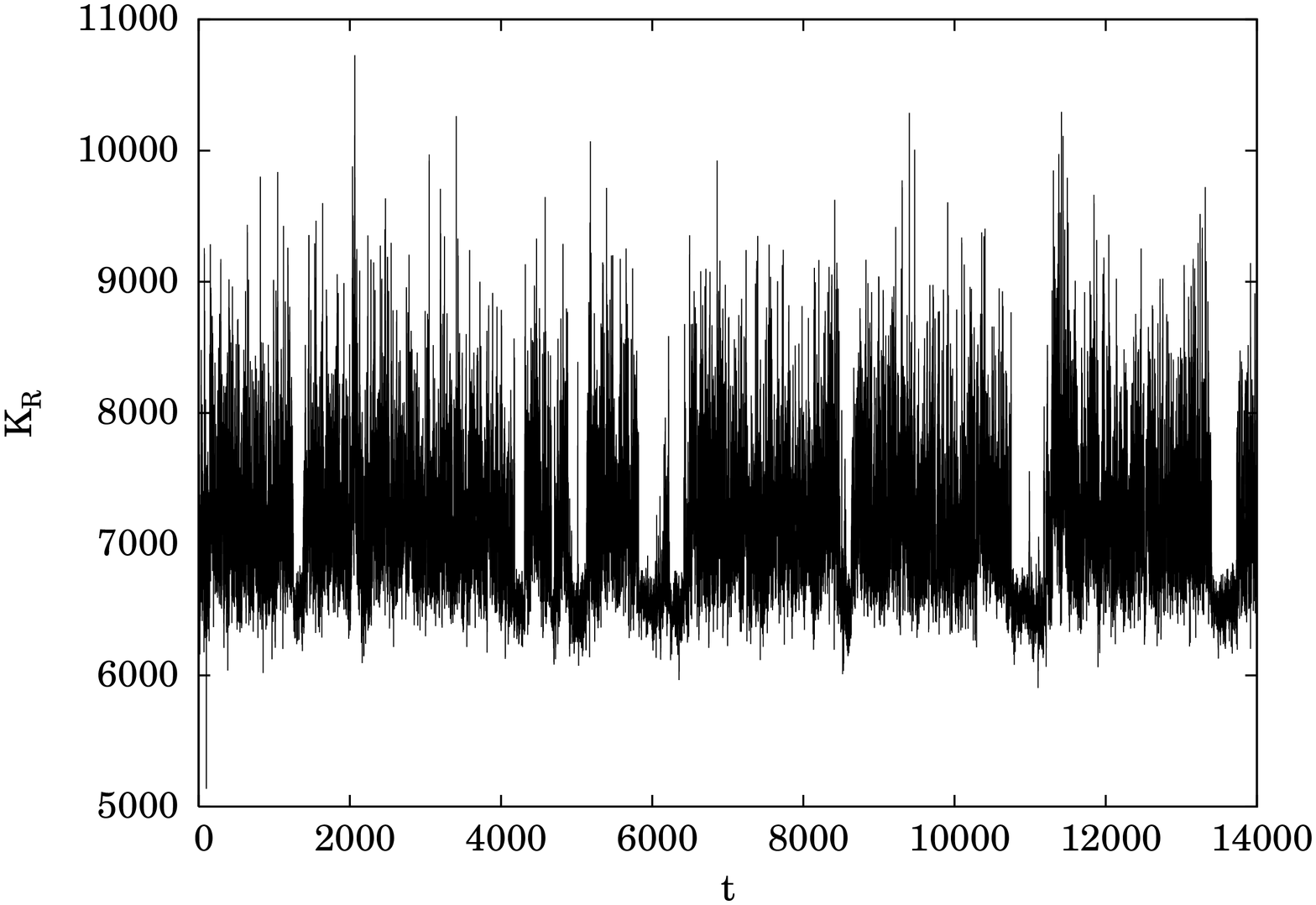}
\end{center}
\caption{$K_R(t)$ for the homopolymer $S^{22}_{\text{h}}$ at $T \approx T_\theta$ (on the
left)  and for the
protein-like sequence $S^{22}_{\text{g}}$ at $T \approx T_f$ (on the right).}
\label{figure_time_series}
\end{figure}

\begin{figure}
\begin{center}
\includegraphics[width=7cm,clip=true,angle=-90]{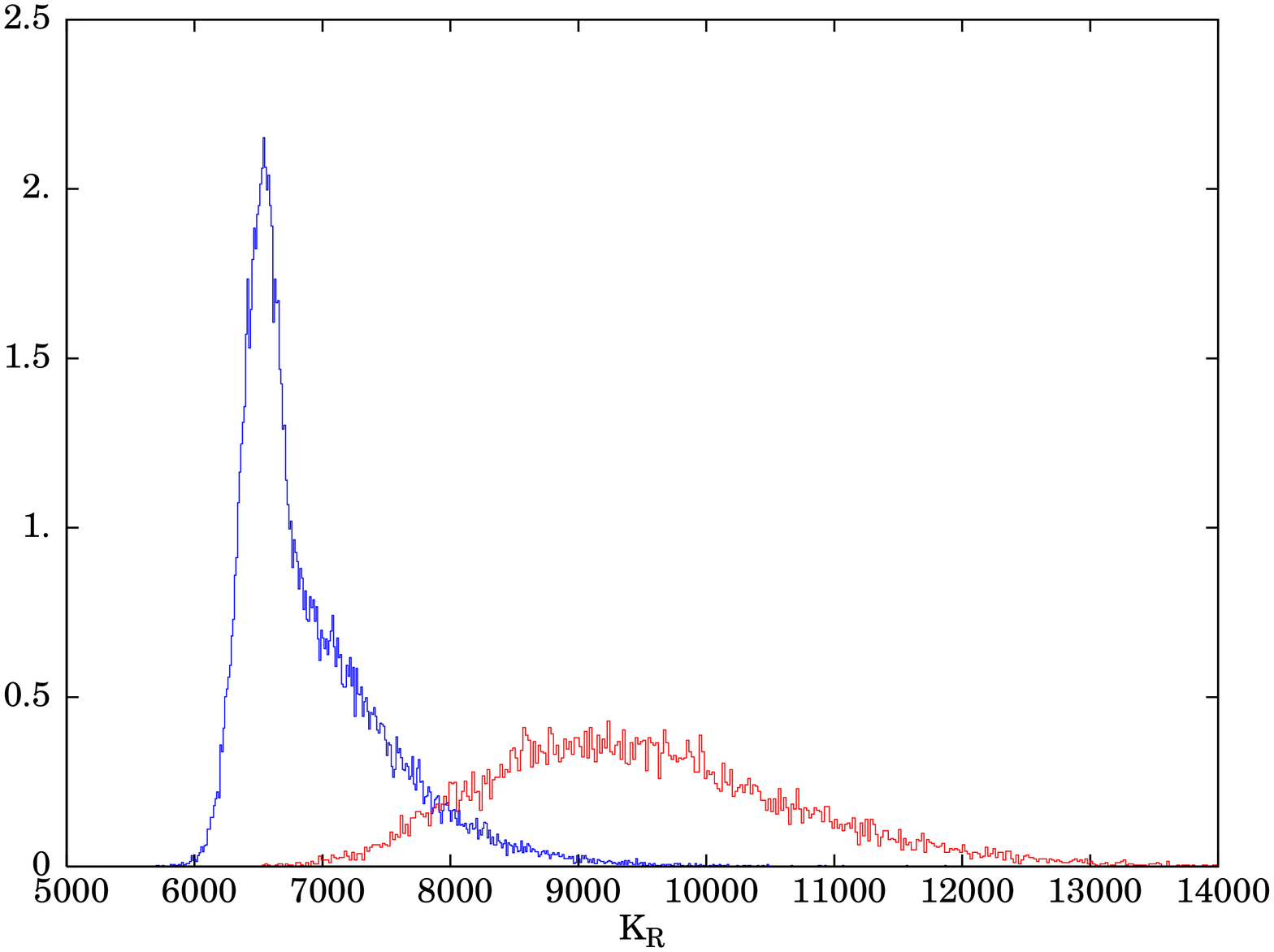}
\end{center}
\caption{(Color online) Normalized distributions of $K_R$ for the homopolymer $S^{22}_{\text{h}}$ at 
$T \approx T_\theta$ (broad histogram) and for the protein-like sequence $S^{22}_{\text{g}}$ 
at $T \simeq 0.68$ (peaked histogram). The values in the vertical axis must be
multiplied by $10^{-7}$. The skewness of the distributions is 0.6 for the
homopolymer case and 1.4 for the protein-like case.}
\label{fig_hist}
\end{figure}

Where does the peak in $\sigma(T)$ near $T_f$ come from? To find clues towards an answer
we may have a closer look at the properties of the curvature of the energy landscape.
Looking at the time series of the curvature $K_R(t)$ for the two sequences 
$S^{22}_{\text{g}}$ and  $S^{22}_{\text{h}}$, i.e., the good folder and the homopolymer
of length 22, respectively, a clear difference shows up when we consider data sampled
close to the transition temperatures, $T_f$ and $T_\theta$, respectively 
(Fig.\ \ref{figure_time_series}). 
In the homopolymer, $K_R(t)$ oscillates in an apparently random fashion around a constant
mean value;
the data reported in Fig.\ \ref{figure_time_series} (left panel) have been taken at $T\approx
T_\theta$, but---in the case of the homopolymer---they are not qualitatively 
different from those taken at any other
temperature.
On the contrary, the curvature signal of the protein-like sequence near $T_f$ (right panel in
Fig.\ \ref{figure_time_series}) is much more
structured: there are several time windows where the curvature oscillates around a mean
value which is considerably lower than the global one, and in these windows also the
amplitude of the fluctuations is smaller. 
\begin{figure}
\begin{center}
\includegraphics[width=7cm,clip=true,angle=-90]{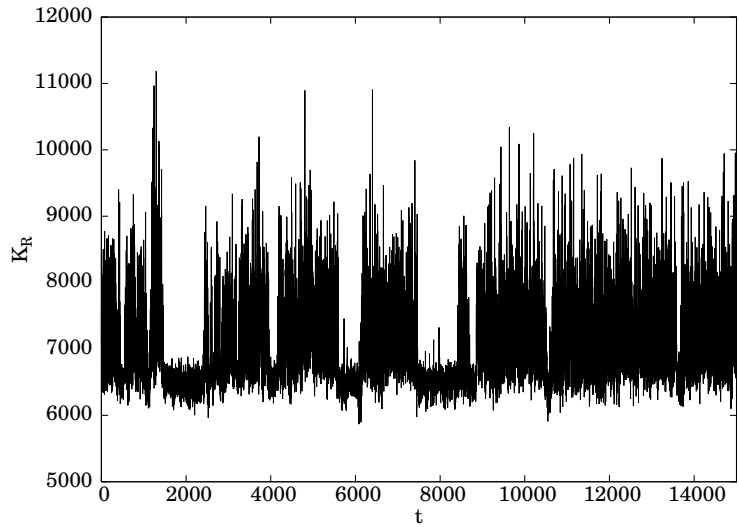}
\end{center}
\caption{$K_R(t)$ for the 
protein-like sequence $S^{22}_{\text{g}}$ at $T \simeq 0.68$, i.e., at the temperature of the peak
in the relative curvature fluctuations $\sigma(T)$.}
\label{fig_peak_series}
\end{figure}
This effect is even more visible at the temperaure of the peak in the curvature fluctuations
$\sigma(T)$ (which is slightly larger than the estimated $T_f$, although within the estimated
errorbar), as shown in 
Fig.\ \ref{fig_peak_series}, and leads to an asymmetric distribution of the
values of $K_R$, resulting in ``anomalously large'' fluctuations which are at the origin
of the peak of $\sigma(T)$ close to $T_f$. Histograms of the distributions of $K_R$ for the
homopolymer at $T \approx T_\theta$ and for the sequence $S^{22}_{\text{g}}$ at the temperature of
the peak in the fluctuations, $T\simeq 0.68$, are shown in Fig.\ \ref{fig_hist}. Also the
distribution of curvatures for the homopolymer is asymmetric, but the asymmetry is considerably
lower than in the good folder case.

\begin{figure}
\begin{center}
\includegraphics[width=7cm,clip=true,angle=-90]{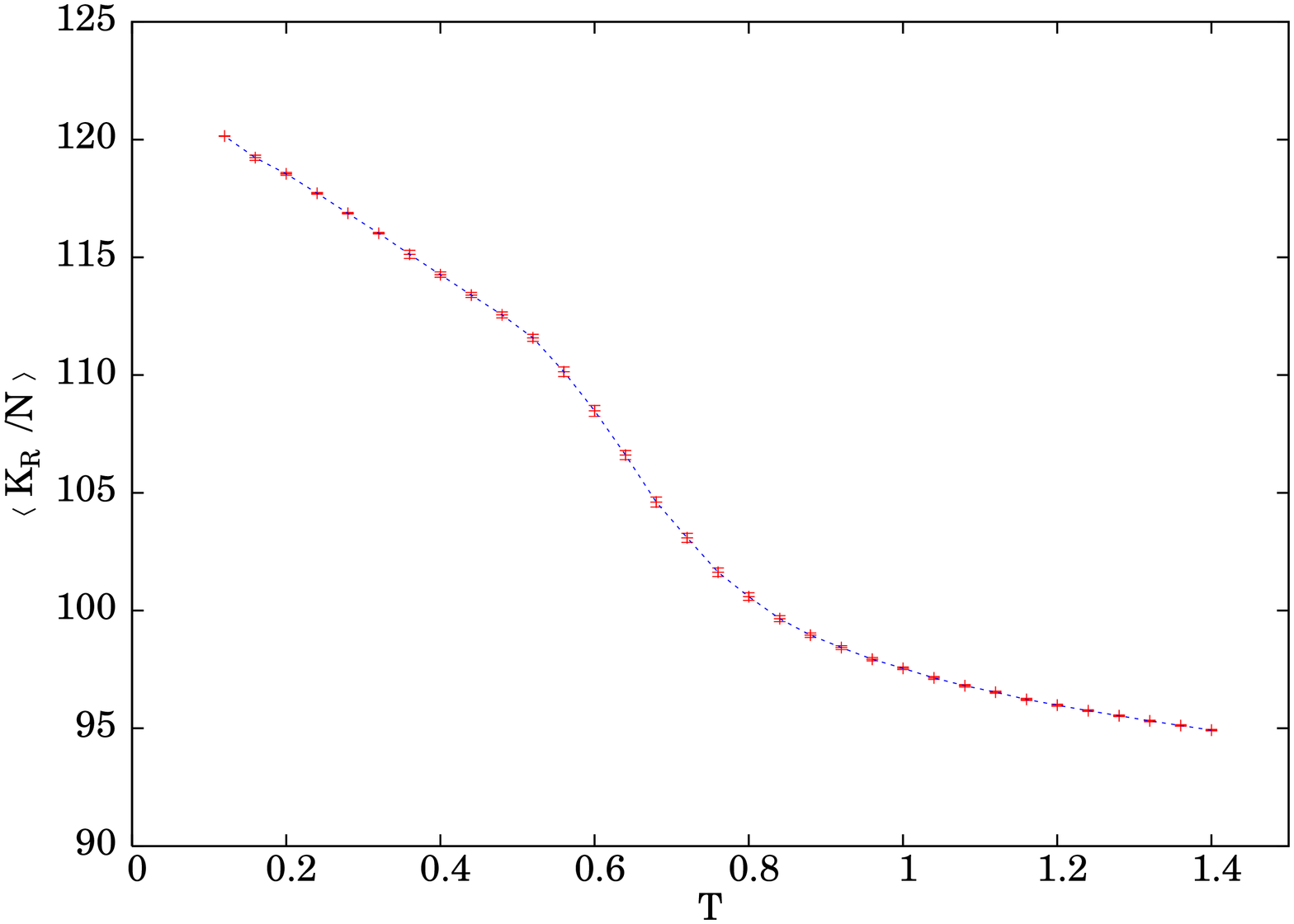}
\end{center}
\caption{(Color online) Average curvature (per degree of freedom) $\langle K_R \rangle /N$ for the 
protein-like sequence $S^{22}_{\text{g}}$ as a function of $T$.}
\label{fig_average_curv}
\end{figure}

\begin{figure}
\begin{center}
\includegraphics[width=7cm,clip=true,angle=-90]{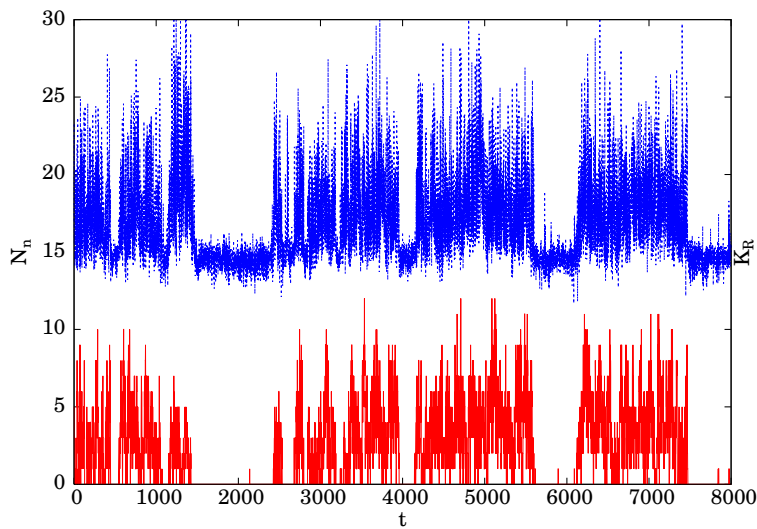}
\end{center}
\caption{(Color online) Synopsis of the instantaneous values of the 
curvature $K_R$ (upper curve) and of the number of
native contacts $N_n$ (lower curve) 
for the protein-like sequence $S^{22}_{\text{g}}$ in a simulation run
at $T \simeq 0.68$.}
\label{fig_curv_nat}
\end{figure}

We interpret these data as direct indications of the presence of two macroregions in the energy
landscape, one corresponding to the native state and the other---charatcterized by a
smaller average curvature---corresponding to the unfolded state. Then the dynamics is
effectively two-state: close to $T_f$, the system often jumps between the two basins and
this explains the behavior of the observed time series of $K_R$. 
This interpretation is supported also by the following observations. First, the average curvature
is a decreasing function of $T$ (Fig.\ \ref{fig_average_curv}); second, comparing the
instantaneous values of the curvature and of the number of native contacts (defined 
in Eq.\ \ref{mio_ij}) we clearly see that the time windows where the curvature is lower correspond
to zero native contacts, thus indicating that the system is in the unfolded state 
(Fig.\ \ref{fig_curv_nat}).

This suggests that an
effective description of this system in terms of a coarse-grained energy landscape should
be possible: work is in progress along this direction \cite{model_1d}.

\section{Concluding remarks}

Studying six different sequences of a minimalistic model of a protein,  we have shown that a
geometric quantity which measures the amplitude of the  curvature fluctuations of the energy
landscape, $\sigma$, when plotted as a function of the temperature $T$  shows a dramatically
different behavior when the system undergoes a folding transition with respect to when only a
hydrophobic collapse is present; $\sigma(T)$ can thus be used to mark the folding transition
and to identify good folders, within the model considered here. 

It must be stressed that no knowledge of the native state is necessary to define $\sigma$,
and that it can be computed with the same computational effort needed to obtain the specific
heat and other thermodynamic observables. This means that using  e.g.\ reweighting histogram
techniques one can reliably estimate the behavior of $\sigma$ as a function of $T$ using few
simulation runs at properly chosen temperatures, if not a single one. Hence, if tested
successfully on other, maybe more refined models of proteins, the calculation of the
curvature fluctuations might prove a useful tool in the search of protein-like sequences.
Preliminary results on more refined minimalistic models \cite{noi_cecilia} completely confirm
the scenario presented here.

In case numerical evidence
accumulates and/or general theoretical arguments become available in favor of
our suggestion that the peak in the curvature fluctuations is a generic marker
of a protein-like behavior, then this method could provide, for
example, a fast and reasonably cheap numerical tool to make a preliminary discrimination 
between sequences that are likely to be protein-like and the others. This may be useful not
only when trying to understand the general features of energy landscapes of minimalistic
models, but also for tasks of more applicative interest like assessing the relevance of point
mutations in a sequence.

We have shown indications that the presence of a peak in the curvature
fluctuations when the system undergoes a folding is a
consequence of the effective two-state dynamics of this system close to the folding
transition. On the basis of the sole results presented here such an interpretation can not be
considered as definitive; nonetheless, it is confirmed by preliminary results on an effective model
of the system studied here, which will be discussed in a forthcoming paper \cite{model_1d}.
It must also be noticed that higher curvature fluctuations imply in general a
higher degree of instability of the dynamics (see \cite{physrep}), as expected near $T_f$
where the system has essentially the same probabilty of being in two very different
states, folded or unfolded. A deeper investigation of the instability properties of the
dynamics close to the folding transition would probably yield more interesting
information.

Apart from the possible applications of $\sigma(T)$ as a diagnostic tool, and even more as giving
indications about the global structure of the energy landscape, the results we
have presented here may 
also be interesting because open a connection between the folding transition and
symmetry-breaking phase transition. The behavior of $\sigma(T)$ observed here for the
good folders is remarkably close to that exhibited by finite systems undergoing a
symmetry-breaking phase transition in the thermodynamic limit \cite{TdF_geometry}, while
the case of the homopolymers is similar to that of a Bere\v{z}inskij-Kosterlitz-Thouless 
transition. 
This suggests that the folding of a proteinlike heteropolymer does share some features of
``true'' symmetry-breaking phase transitions---at least those features that show up already in
finite systems---although no singularity in the thermodynamic limit occurs, because
proteins are intrinsically finite objects \cite{Dill, CecconiBurioni}. The behavior of
$\sigma(T)$ in systems with thermodynamic phase transitions has been interpreted in terms
of topological changes of the manifolds where the dynamics of the system ``lives''
\cite{cccp,franz} (see also \cite{physrep,marco_book} and \cite{kastner_rmp} for a review). 
Work is in progress to understand 
whether such a topological interpretation can be extended also to the case of the
folding transition, although it has no infinite-system,  thermodynamic limit
counterpart.

\acknowledgments

We thank Lorenzo Bongini and Aldo Rampioni for useful discussions and
suggestions. This work is part of the EC (FP6-NEST) project {\em Emergent
organisation in complex biomolecular systems (EMBIO)} (EC contract n.\ 012835)
whose financial support is acknowledged.

\end{document}